\documentclass[a4paper,10pt,twocolumn,accepted=2020-06-29]{quantumarticle}

\pdfoutput=1

\usepackage[utf8]{inputenc}
\usepackage[english]{babel}
\usepackage[T1]{fontenc}
\usepackage{amsmath}
\usepackage{hyperref}
\usepackage{tikz}

\usepackage{float}
\usepackage{mathtools}
\usepackage{soul}
\usepackage{amsmath,amssymb}
\usepackage{tabu}
\usepackage{booktabs}
\usepackage{array}
\usepackage{multirow}
\usepackage{nicefrac}
\usepackage{graphicx}
\usepackage[super, comma, sort&compress]{natbib}
\usepackage{xspace}
\usepackage[export]{adjustbox}[2011/08/13]
\usepackage{pifont}
\usepackage[a4paper, total={7.5in, 10.5in}]{geometry}
\usepackage{hyperref}

\def\cz{\ensuremath{\text{CZ}}\xspace}
\newcommand{\ket}[1]{\ensuremath{|#1\mkern-1mu\rangle}}
\newcommand{\dyad}[1]{\ensuremath{|#1\rangle\mkern-3mu\langle #1|}}

\newcommand{\ch}[1]{{\color{black}#1}}

\newcommand{\jer}[1]{{\color{black}#1}}

\newcommand{\xmark}{\ding{55}}%
\newcommand{\cmark}{\checkmark}

\newcolumntype{P}[1]{>{\centering\arraybackslash}p{#1}}

\begin{document}




\title{Mapping graph state orbits under local complementation}

\author{Jeremy C. Adcock}
\email{jeremy.adcock@bristol.ac.uk}
\affiliation{Quantum Engineering Technology (QET) Labs, H. H. Wills Physics Laboratory \& Department of Electrical \& Electronic Engineering, University of Bristol, Merchant Venturers Building, Woodland Road, Bristol BS8 1UB, UK}
\orcid{0000-0002-8923-7180}

\author{Sam Morley-Short}
\affiliation{Quantum Engineering Technology (QET) Labs, H. H. Wills Physics Laboratory \& Department of Electrical \& Electronic Engineering, University of Bristol, Merchant Venturers Building, Woodland Road, Bristol BS8 1UB, UK}
\orcid{0000-0002-4445-734X}

\author{Axel Dahlberg}
\affiliation{QuTech - TU Delft, Lorentzweg 1, 2628CJ Delft, The Netherlands}
\orcid{0000-0003-2479-7424}

\author{Joshua W. Silverstone}
\affiliation{Quantum Engineering Technology (QET) Labs, H. H. Wills Physics Laboratory \& Department of Electrical \& Electronic Engineering, University of Bristol, Merchant Venturers Building, Woodland Road, Bristol BS8 1UB, UK}
\orcid{0000-0002-3429-4890}

\maketitle

\begin{abstract}
        Graph states, and the entanglement they posses, are central to modern quantum computing and communications architectures. 
    Local complementation---the graph operation that links all local-Clifford equivalent graph states---allows us to classify all stabiliser states by their entanglement.
    Here, we study the structure of the orbits generated by local complementation, mapping them up to 9 qubits and revealing a rich hidden structure. 
    \ch{We provide programs to compute these orbits, along with our data for each of the $587$ orbits up to $9$ qubits and a means to visualise them.
    We find direct links between the connectivity of certain orbits with the entanglement properties of their component graph states.
    Furthermore, we observe the correlations between graph-theoretical orbit properties, such as diameter and colourability, with Schmidt measure and preparation complexity and suggest potential applications.}
    It is well known that graph theory and quantum entanglement have strong interplay---our exploration deepens this relationship, providing new tools with which to probe the nature of entanglement.
\end{abstract}

\vspace{0.5cm}
\section{Introduction}

Graph states provide a language of entanglement between qubits and are at the core of modern quantum computing and communication architectures across all qubit platforms\cite{raussendorf2009measurement, veldhorst2017silicon, lekitsch2017blueprint, alexander2016one, asavanant2019time, barends2014superconducting, markham2008graph}.
Graph states are a subset of stabiliser states. Some stabiliser states are local-Clifford (LC) and every stabiliser state is local-Clifford equivalent to at least one graph state. Graph states which are LC equivalent are related by repeated application of a simple graph operation, local complementation\cite{hein2004multiparty, van2004graphical}. Hence all sets of LC-equivalent stabiliser states can be completely described by sets, or `classes', of graphs. \ch{Note that states which are LC equivalent are also local unitary equivalent up to at most 27 qubits\cite{ji2007lu}, with a lower bound of 8 qubits\cite{cabello2009entanglement}.}
Since local operations cannot change the type of entanglement a state possesses, graph states provide a way to classify all stabiliser states by the entanglement they posses.

Graph state entanglement is well studied \cite{hein2004multiparty, van2004graphical, hein2006entanglement, dahlberg2018transforming, dahlberg2019complexity}, with each of the ${\sim}1.6\times10^{12}$ non-isomorphic graph states up to 12 qubits classified into ${\sim}1.3\times 10^6$ LC-inequivalent classes\cite{danielsen2006classification, cabello2011optimal}. 
There is a polynomial time algorithm to compute the LC unitary relating two graph states (if there is one)\cite{bouchet1991efficient, van2004efficient}. 
In contrast, the problem of determining if a target graph state can be generated from an input graph state using LC operations, local Pauli measurements and classical communication (LC+LPM+CC) is \textsc{np}-complete for both labelled \cite{dahlberg2018transforming, dahlberg2018transform} and unlabelled graphs\cite{dahlberg2019complexity}. 
It is also known that counting single-qubit LC-equivalent graph states is $\#\textsc{p}$-complete\cite{dahlberg2019counting}.
\ch{Due to this hardness, exploration have been limited to $n\leq12$ qubits.
Ref.~\citenum{cabello2011optimal} supplies tables containing information on every entanglement orbit for $n\leq12$ as supplementary material.
This includes a canonical member graph state for each orbit, as well as quantities relating to that state, and other classifying information. For example, the minimum edge number of a graph state in the class is given, along with bounds on its Schmidt measure and the number of graph states in the class.}

\begin{figure*}[t!]
\centering

\captionsetup{width=1.0\textwidth}
\includegraphics[width=1.0\textwidth,center]{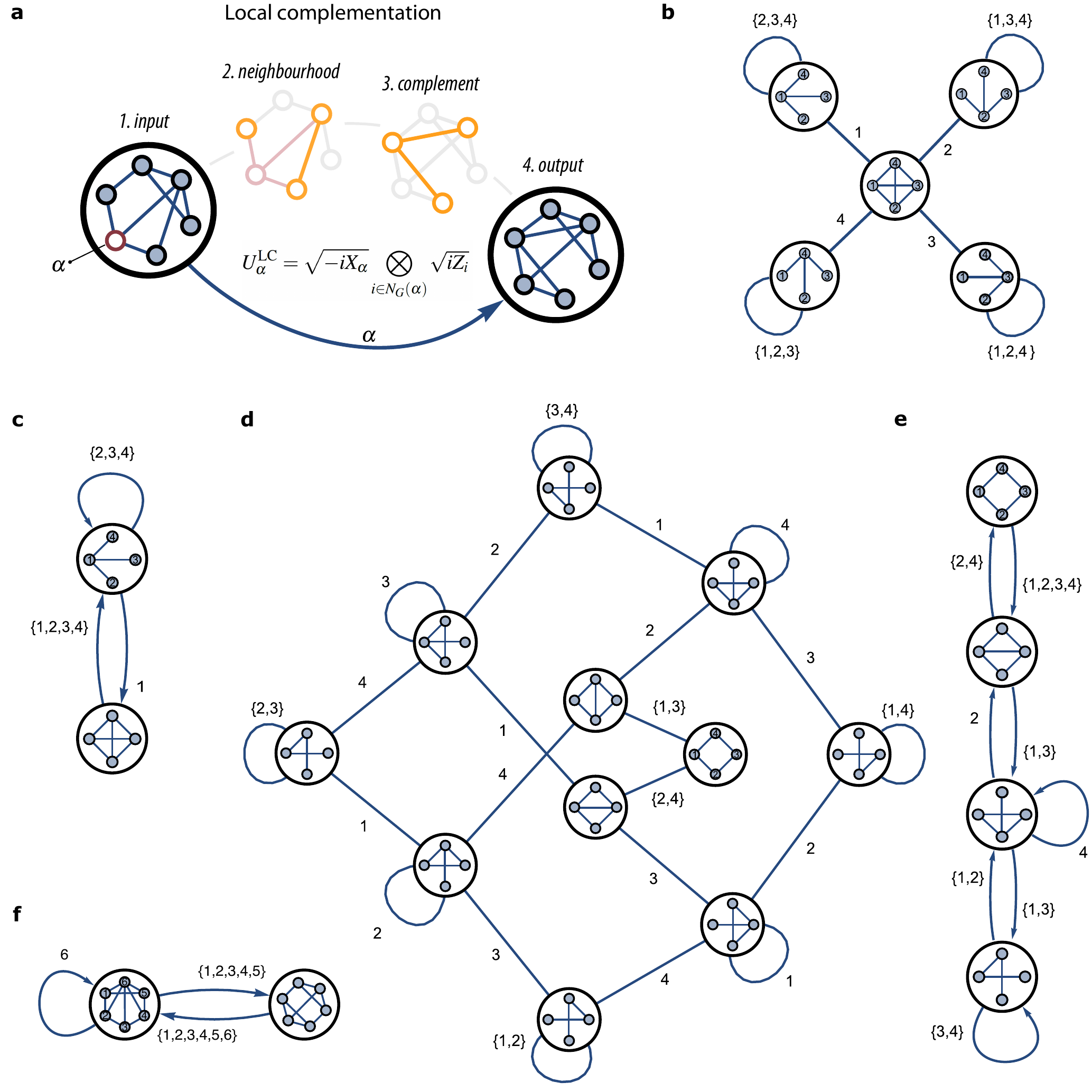}

\caption{local complementation and the orbits it induces. Orbit edges are labelled with the vertex that undergoes local complementation. \textbf{a.} A guide to local complementation. The neighbourhood of qubit $\alpha$ is complemented to yield the output graph. \textbf{b.} The orbit $L_3$ (GHZ entanglement of four qubits). \ch{Here, $L_i$ and $C_i$ denote the orbit induced by local complementation for entanglement class $i$.} \textbf{c.} The orbit $C_3$, where isomorphic graph states are considered equal. \textbf{d.} The orbit $L_4$ (cluster state entanglement of four qubits). This is one of three equivalent orbits, which together contain every isomorphism of the contained graph states. \textbf{e.} The orbit $C_4$. \textbf{f.} The orbit $C_{19}$. Graph state vertices are labelled descending clockwise from noon (see \textbf{b}). We use directed edges when drawing $C_i$ orbits as only one isomorphism of the graph states can be drawn on an orbit.
}
\label{img:line4ghz4}
\end{figure*}

\jer{Recently we showed that local complementation can be used to generate graph states more efficiently\cite{adcock2018hard}.}
However, little is known about the structure of the orbits that are generated by local complementation.
These orbits are themselves graphs, in which each orbit vertex represents a graph state and edges between them are induced by local complementation of different graph state vertices (see Fig.~\ref{img:line4ghz4}). 
Here, we refer to the object that links graphs via local comeplemtation as their `orbit', and we refer to those component graphs as `graph states'.
These orbits, which are wildly complex, give a fresh perspective for the study of stabiliser entanglement and graph states, while providing new tools for optimising quantum protocols.

\ch{Where previous work has `catalogued' each class of graph states and provided a set of graphs for each class, in this work we focus on understanding the structure of how each graph state is related to the others via local complementation, by 'mapping' the space in which they live.}
To do so we generate the orbit of each of the 587 entanglement classes up to $n \leq 9$ qubits.
\ch{We also provide\cite{gsc} `graph state compass' a new tool to generate the orbit generated by local complementation given an input graph state, along with all of the data generated in this study and the code used to generate the plots found in this manuscript\cite{adcock2019graphorbitsonline}.}
We compute graph-theoretical properties of these orbits and link these to properties of the member graph states, while observing strong correlations between orbit complexity and known entanglement metrics.
We also identify promising applications of local complementation in both quantum secret sharing and compilation of measurement-based protocols.
\ch{By mapping these orbits we expose the exquisite structure of graph state orbits and present them as promising avenues for further study.}

\section{Graph state orbits}

Graph states are quantum states with a one-to-one correspondence to mathematical graphs\cite{hein2004multiparty, van2004graphical}.
A graph, $G=(V,E)$, is a combinatoric object defined by a set of edges $E$ between a set of vertices $V$.
The corresponding graph state is written:
\begin{equation}
\ket{G} = \prod_{(i,j) \in E} \text{\text{\text{CZ}}}_{ij} \ket{\text{+}}^{\otimes |V|}.
\end{equation} 
Here, $\ket{+} = (\ket{0} + \ket{1})/\sqrt{2}$ and $ \cz = \dyad{00} + \dyad{01} + \dyad{10} - \dyad{11}$.
Connected $n$-vertex graphs have genuine $n$-partite entanglement.

Remarkably, graph states can be LC-equivalent, despite having different constructions via nonlocal Controlled-$Z$ (CZ) gates\cite{hein2004multiparty, van2004graphical}.
Specifically, graphs are LC-equivalent if and only if they can be transformed into one another by successive applications of local complementation.

Local complementation of a vertex $\alpha$, $\text{LC}_{\alpha}$, applied to a graph, $G(V,E)$, acts to complement the neighbourhood of the vertex $\alpha$.
That is, in the neighbourhood of $\alpha$, it removes edges if they are present, and adds any edges are missing (see Fig.~\ref{img:line4ghz4}a).
More formally:
\begin{equation}
    \text{LC}_{\alpha} (G(V,E)):\rightarrow G(V,E'),
\end{equation}
where
\begin{equation}
E' =  E \cup K_{N_G(\alpha)} - E \cap K_{N_G(\alpha)} =E \Delta K_{N_G(\alpha)}.
\end{equation}
Here, $K_{N_G(\alpha)}$ is the set of edges of the complete graph on the vertex set $N_G(\alpha)$, the neighbourhood of $\alpha$, and, $\Delta$ is the symmetric difference.
On graph states, the following local  unitary implements local complementation\cite{hein2004multiparty, van2004graphical}:
\begin{equation}
U^{\mathrm{LC}}_{\alpha} = \sqrt{-iX_{\alpha}} \bigotimes_{i \in {N_G(\alpha)}} \sqrt{i Z_{i}}
\end{equation}
Where $U^{\mathrm{LC}}_{\alpha}\ket{G} = \ket{\mathrm{LC}_\alpha (G)}$. Repeated application of local complementation is guaranteed to hit every member of a entanglement class of LC-equivalent graph states, given any member of that class as a starting point\cite{hein2004multiparty, van2004graphical}.
This defines graph (and therefore stabiliser) entanglement classes, each with their own orbit under local complementation.
\ch{Though these classes have been catalogued\cite{cabello2011optimal} up to $n=12$, to our knowledge the structure of their orbits as not yet been investigated.}

All $n$-vertex graphs can be locally complemented in $n$ different ways, generating up to $n$ different graphs.
Each of these can be locally complemented further, generating up to $n - 1$ new graphs (local complementation is self inverse). 
We can repeatedly local complement graphs until we find no new ones, concluding that all graphs in the class have been found.
By performing every local complementation on every graph in the class, the orbit is mapped (see Section \ref{sec:ge}). 
We will denote these orbits $L_i$ for entanglement class $i$, canonically indexed as in ref.~\citenum{cabello2011optimal}.

This orbit is itself naturally represented as a graph---its vertices are graph states and the edges that link them are local complementations on the graph state's vertices (see Fig.~\ref{img:line4ghz4}).
Edges of the orbits are labelled with a vertex index indicating which local complementation links the two graph states on the orbit vertices. Since local complementation is self-inverse, these edges are undirected. 
Some simple examples of orbits are shown in Figs.~\ref{img:line4ghz4}b,d.

\subsection{A quantum Rubik's cube}

Local complementation orbits have an entertaining analogy with the popular puzzle toy, the Rubik's cube.
Each face of a Rubik's cube is a different colour, which is itself separated into $3\times3 = 9$ individual squares.
This is the cube's solved state.
The toy has $6$ basic moves, which rotate the different faces of the cube by $90^\circ$. 
By applying these six moves in a random combination, a random state of the cube is generated.
The challenge is then to return the cube to its solved state.
For a mathematician, the challenge is to understand the cube's symmetry, and solve it in the general case.

Using about one billion seconds (35 years) of CPU time, the Rubik's cube Cayley graph---the orbit of the states of the cube--has been computed\cite{rokicki2014diameter}.
Indeed, a Rubik's cube has ${\sim}4.3 \times 10^{19}$ states and its orbit has diameter $26$.
That is, any Rubik's cube can always be solved in $26$ $90^\circ$ moves or less ($20$ moves if both $90^\circ$ and $180^\circ$ rotations are allowed).
`Cubers', as Rubik's cube aficionados are known, call 26 `god's number'.

In our analogy, the many states of the toy are our graph states, and rotating the different faces of the cube corresponds to local complementation of different graph vertices.
As evidenced by the ratio of its cardinality to its diameter (${\sim} 10^{18}$), the orbit of the Rubik's cube is highly dense (though each vertex only has six edges).
Each of the ${\sim}1.3$ million entanglement classes of $12$ qubits has its own unique orbit---each of them is another Rubik's cube (with $12$ rather than $6$ moves).
Note there are factorially many entanglement classes as $n$ is increased.
God's number (the orbit diameter) for local complementation orbits depends on the class. 
Using about a week of CPU time on a standard desktop computer, we compute the diameter of local complementation is maximally 9 for 9-qubit graph states.
That is, any two LC-equivalent graph states are at most 9 local complementations distant from one another).

\subsection{Isomorphic graph states}
\label{sec:giilco}

Graphs which are identical under relabelling of their vertices are said to be \emph{isomorphic}.
Graph states which are isomorphic share the same variety of entanglement.
This is an important feature for the implementation of protocols where qubit relabelling is non-trivial---this includes most quantum information processing and communication scenarios.
Here we consider both cases.
We denote orbits $C_i$ when isomorphic graphs are considered equal (unlabelled graph states), and $L_i$ otherwise (labelled graph states).
\ch{By examining our dataset, we observe that }$C_i$ contain on average $\nicefrac{1}{8} $ as many graph states as their partner $L_i$ orbits for $n<9$ qubits.
This greatly reduces the computational resources needed to map and analyse them.
\ch{We note that all $C_i$ are subgraphs of $L_i$ for all $i$.
This subgraph is formed by merging all orbit vertices corresponding to isomorphic graph states.}
This can be seen by observing that isomorphic graph states have isomorphic neighbourhoods in $L_i$.

\ch{We find that there are typically more than one \ch{$L_i$} orbit (for fixed $i$), as most \ch{$C_i$} orbits do not contain every isomorphism of it's member graph states (e.g.~Fig.~\ref{img:line4ghz4}d)---the entanglement possessed is distributed in different ways between the parties.
These equivalent orbits are themselves isomorphic, and together the set of $L_i$ (for fixed $i$ orbits) contains every isomorphism of their component graph states.}
For example, there are three equivalent orbits of $L_4$ (one of which is shown in Fig.~\ref{img:line4ghz4}), each containing different isomorphisms of their component graph states.
Some entanglement classes have only one $L_i$ orbit, which contains every isomorphism of the graph states. 
For example, the classes which contain the `star' and fully-connected graph states.
These orbits are composed of $|L_i| = n+1$ graph states (vertices) and are themselves a `star' graph (see Section \ref{sec:results} and Fig.~\ref{img:line4ghz4}b).

As in $L_i$ orbits, edges of a $C_i$ orbit are undirected.
However, as a guide to the eye we display directed edges for $C_i$ orbits when those edges are labelled, as this allows the reader to identify which graph vertex undergoes local complementation to reach the output graph (see Fig.\ \ref{img:line4ghz4}c,e,f).

\subsection{Orbit exploration}
\label{sec:ge}

Mapping the orbit of the $i^{\text{th}}$ entanglement class, $L_i$ containing a graph state $\ket{G}$, is a graph exploration problem.
Here, we use an exhaustive breadth-first exploration to traverse the entire orbit, cataloguing each graph state (vertices of the orbit) along with how local complementation links them (edges of the orbit).
We start with a single graph state $G$, taken from ref.~\citenum{cabello2011optimal}, in our catalogue, and perform each possible local complementation on it.
In doing so, we discover up to $n$ new orbit vertices and up to $n$ new orbit edges.
Then we perform every possible local complementation on those output graph states and catalogue the outputs by comparing them to graph states which we have already found.
This is repeated until every local complementation has been performed on every graph state in the catalogue (and no new graph states or edges are found).
%

%








\ch{To map an $n$-qubit orbit, $L_i$, which contains $|L_i|$ graph states requires $O(n|L_i|^2)$ local complementations and graph comparisons.
By `graph comparison', we mean evaluating if two graphs are equal, (or whether they are isomporphic for a $C_i$ orbit). 
Linear savings can be made by noting that local complementation is self inverse, and has no effect when applied to a vertex of degree 1.}

We use this method to explore the $L_i$  for $n\leq 8$ and $C_i$ for  $n\leq 9$, that is, up to graph state entanglement class $i=146$ and $i=586$ respectively.
The largest of these orbits contains $3248$ and $8836$ graph states, respectively.
\textsc{GraphIsomorphism} is a costly routine, belonging to the complexity class \textsc{np}.
Exploration of $C_i$ makes heavy use of \textsc{GraphIsomorphism}, calling it up to $n|C_i|$ times.
However, since $|C_i| \ll |L_i|$, and our graph states are of modest size, exploring $C_i$ up to $9$ qubits required less computational time than exploring $L_i$ up to $8$ qubits.

\ch{In real-world applications, the physical location of qubits is important---isomorphic graph states can not be considered equal.
However, to our knowledge, $C_i$ entanglement classes have not been studied in detail before.
Usually, most isomorphisms of a graph state are not contained within a given $C_i$ orbit.
Hence, knowledge of $C_i$, or at least its members, may be crucial for measurement-based quantum protocols.
}

Local complementing symmetric vertices of an input graph state will result in the same output graph state.
This observation can be used to improve the efficiency of $L_i$ orbit exploration.
The sets of vertices which result in isomorphic graphs under local complementation can be found by computing the automorphism group of each graph state---vertices that are exchanged in an automorphism result in isomorphic graphs.
For example, in the four-vertex ring graph, all vertices are equivalent and so only a single complementation is required, whereas for the four-vertex line graph, there are two non-equivalent vertices, the `inner' and `outer' vertices.
Hence, by computing the automorphism group of each graph state as it is discovered, and only local complementing the reduced subset of graph state vertices that are not equivalent, a saving can be made.
Here, only $\tilde{n}|C_i|^2$ comparisons (and hence calls to \textsc{GraphIsomorphism}) need be made (where $\tilde{n}=|E_i|/|C_i|$ is the mean number of non-symmetric vertices on the graph states of $C_i$.
In practice, the \textsc{AutomorphismGroup} is computed in order to solve \textsc{GraphIsomorphism}\cite{mckay2014practical}.
Hence a linear speedup is achieved.
By examining our set of computed orbits, we find this technique reduces the number of calls to \textsc{GraphIsomorphism} by at least half for $n\leq9$.
%

\section{Results}
\label{sec:results}

We compute a variety of graph properties of $C_i$ orbits of $3$-$7$ qubits and display them in Table~\ref{tab:orbittab}. 
Definitions of these quantities can be found in Appendix Section \ref{sec:defs}.

\makeatletter\onecolumngrid@push\makeatother
\begin{figure*}
    \includegraphics[width=\linewidth]{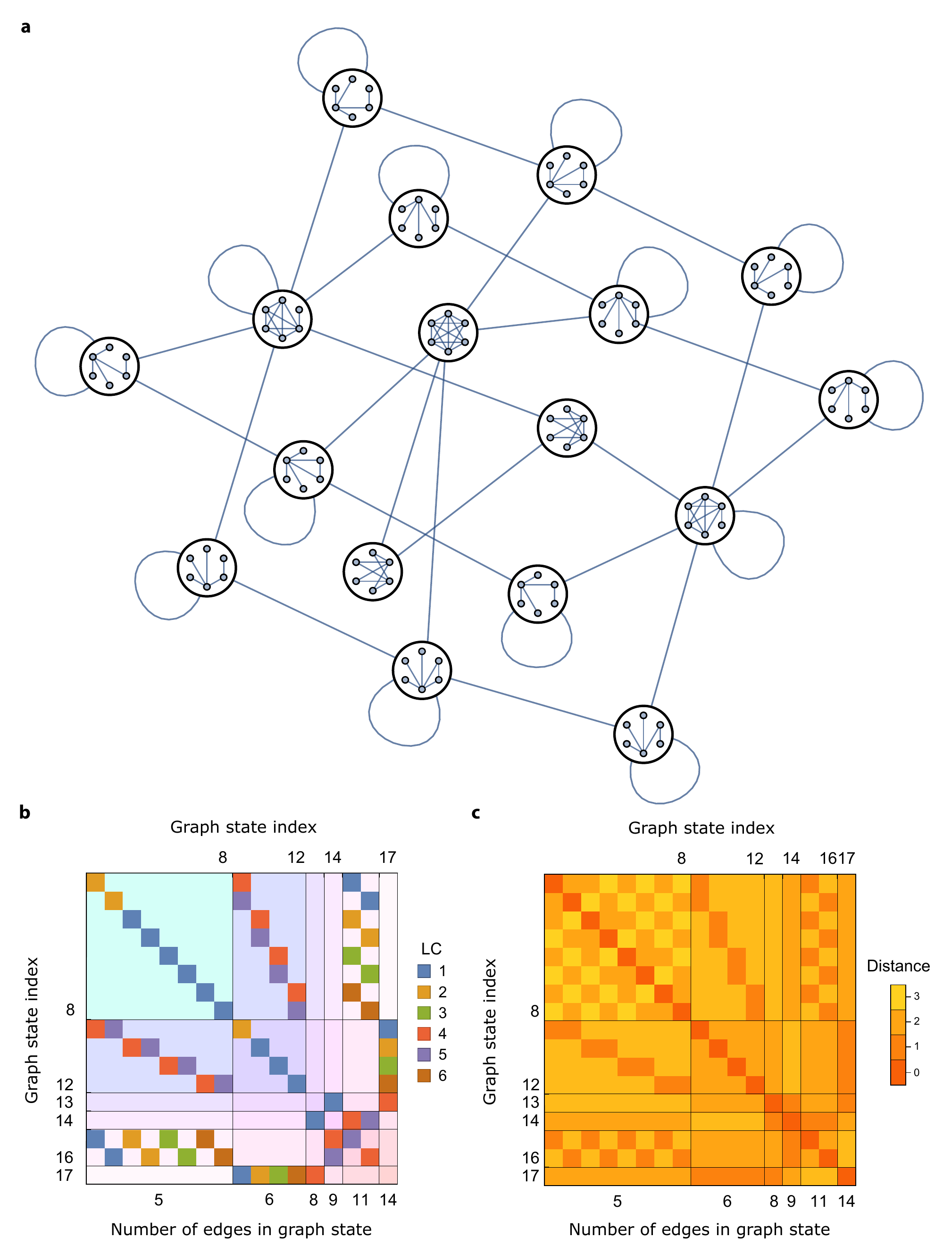}
    \caption{Local complementation orbit $L_{10}$. \textbf{a.} The orbit $L_{10}$. \textbf{b.} The adjacency matrix of $L_{10}$. \textbf{c.} The distance matrix of $L_{10}$. The adjacency matrix of a graph, $A$, has a row and column for each of the graph's vertices. For each edge $(i,j)$ present in graph we write $\Gamma_{ij}=n$, where $n$ is the lowest index of a local complementation that links them. Otherwise $\Gamma_{ij}=0$. Similarly, the distance matrix, $D$, gives the distance between two vertices: $D_{ij}$ is equal to the minimum number of edges that must be traversed to get from vertex $i$ to vertex $j$. Regions of corresponding to isomorphic graph states are demarcated.}
    \label{img:adjmat-6-2}
\end{figure*}
\clearpage
\makeatletter\onecolumngrid@pop\makeatother

For example we display the \ch{Schmidt measure}, $E_S$, which is known to be a useful entanglement monotone for graph states\cite{eisert2001schmidt, hein2004multiparty}, \ch{encoding the strength of error correcting codes built from the state\cite{schlingemann2001quantum}.}
\ch{We also compute the graph state's rank-width\cite{van2007classical, dahlberg2018transforming}, $\mathrm{rwd}(G)$, which plays a fundamental role the complexity of graph state properties: any graph state property which is expressible in so-called monadic second-order logic (a higher-order logical system) can be computed in time $O(f(\mathrm{rwd}(G))|V(G)|^3)$, where $f$ is an exponential function\cite{dahlberg2019complexity}. 
These properties are therefore known as `fixed-parameter tractable', as they are polynomial for graphs with fixed rank-width.
This includes the vertex minor problem, deciding whether a graph can be generated from another with only LC+LPM+CC \jer{operations.}
 It is also known that to be a universal resource for quantum computation, lattice graph states must have unbounded rank-width as they increase in size\cite{van2007classical}.
 The rank-width of every graph state with $n\leq9$ qubits is available in our online resource\cite{adcock2019graphorbitsonline}.
We also provide a host of other graph theoretical properties of the orbit and their graph states, for example their chromatic number, their diameter, and the size of their automorphism group.}

\begin{table}[b!]
\centering
\begin{adjustbox}{center}

\begin{tabular}{P{1.8cm}	P{3.2cm}   P{2cm}}
\ch{Orbit type}	& 	\ch{Correlation coefficient}	    &  \ch{Value}			\\
\noalign{\vskip 1mm}    
\hline
\hline
\noalign{\vskip 1mm}

$C_i$	    	&	$r(\mathrm{max}(d_{jk}^{C_i}), \; |C_i|)$   &	$0.62 \pm 0.03$ \\
$C_i$	    	&	$r(\mathrm{max}(d_{jk}^{C_i}), \; E_S)$	    &	$0.77 \pm 0.02$ \\
$C_i$	    	&	$r(\chi_{C_i}, \; E_S)$				&	$0.67 \pm 0.02$		\\
$C_i$		    &	$r(\chi_{C_i}^e, \; E_S)$			&	$0.81 \pm 0.04$		\\
$C_i$		    &	$r(\chi_{C_i}, \; \chi_{g}^e)$		&	$0.032\pm 0.04$		\\
\noalign{\vskip 2mm}
$L_i$	    	&	$r(\mathrm{max}(d_{jk}^{L_i}), \; |L_i|)$	&	$0.60 \pm 0.05$ 	\\
$L_i$	    	&	$r(\mathrm{max}(d_{jk}^{L_i}), \; E_S)$	    &	$0.93\pm 0.02$  \\
$L_i$	    	&	$r(\chi_{L_i}, \; E_S)$				&	$0.70 \pm 0.05$  	\\
$L_i$	    	&	$r(\chi_{L_i}^e, \; E_S)$			&	$0.44 \pm 0.11$ 	\\
$L_i$	    	&	$r(\chi_{L_i}, \; \chi_{g}^e)$		&	$-0.09 \pm 0.09$ 	\\
\noalign{\vskip 2mm}
--	        	&	$r(E_S, \; \mathrm{rwd})$			&	$0.62 \pm 0.03$		\\
--	        	&	$r(E_S, \; |e|) $					&	$ 0.78 \pm 0.02$	\\
--	        	&	$r(E_S, \; \chi_{g}^e) $			&	$-0.17 \pm 0.02$	
\end{tabular}
\end{adjustbox}
\caption{\ch{Summary of the correlations observed. Here, $E_S$ is the \ch{Schmidt measure} of the orbit, $d_{jk}$ are the distances between graph states in the orbit,  $\chi$ is the chromatic number of the orbit, $\chi^e$ is the chromatic index of the orbit, $\chi_{g}^e$ is the lowest chromatic index of a graph state in the orbit, $\mathrm{rwd}$ is the rank-width of the orbits' graph states and $|e|$ is the minimum number edges of any graph state in the orbit. `--' indicates that the tested property depends only on the set of graph states in the orbit, and not the orbit structure.}}
\label{tab-data}
\end{table}

As per the canonical indexing of graph state entanglement classes, we list the minimum degree of each orbit: the smallest number of edges of any of the orbit's graph states.
Using only CZ gates, this is the minimum number gates needed to produce the entanglement class from $\ket{+}$, .
We also provide the graph state's minimum chromatic index (minimal edge colouring number), which corresponds to the minimum number of time steps required to generate a state in that entanglement class using only CZs\cite{cabello2011optimal}.
\jer{Here, we assume CZs can be performed between each qubit arbitrarily, and note that interspersing CZs with LCs can reduce the number of CZs required.}

We find correlations between orbit parameters and compute their Pearson correlation coefficients, $-1<r(x, \; y)<1$, for orbit parameters $x,y$.
Here, $r = 1$ implies there is exact linear correlation in the data, $r=-1$ indicates an exact negative linear correlation, and $r=0$ implies no linear correlation whatsoever.
To quantify entanglement of graph states, we examine the Schmidt measure, a well-studied entanglement monotone with many convenient relationships to graph states\cite{eisert2001schmidt, hein2004multiparty, danielsen2006classification}.
\jer{For example, it is known that and any graph state that corresponds to a \emph{maximum distance separable} (MDS) error correcting code must have Schmidt measure at least\cite{hein2004multiparty} $|V|/2$.
MDS codes are optimal error correcting codes in that they are able to correct the greatest number of errors for a given number of logical and physical qubits---that is, they saturate the \emph{singleton bound}.}

We observe that the graph state \ch{Schmidt measure}, $E_S$, correlates strongly with orbit diameter ($r(\mathrm{max}(d_{jk}), \; E_S) = \{0.77 \pm 0.02, 0.93\pm 0.02\}$), where the first value is for $C_i$ and the second is for $L_i$.
\ch{Interestingly, orbit diameter correlates more significantly with Schmidt rank than with orbit size ($r(\mathrm{max}(d_{jk}), \; |O_i|) = \{0.62\pm 0.03, 0.60 \pm 0.05\}$).
This indicates that more entangled states are likely to have large, sparse orbits.}
Here, if the \ch{Schmidt measure} is not known, we take the average value of the bounds, which are rarely loose. 
Furthermore, orbit chromatic number, \ch{$\chi_{i}$}, and \ch{Schmidt measure}, \ch{$E_S$} have high correlation coefficients of $r(\chi_{i}, \; E_S) = \{0.67 \pm 0.02,  0.70 \pm 0.05\}$.
Interestingly, orbit chromatic number, \ch{$\chi_{i}$}, does not correlate with minimum graph state chromatic index, \ch{$\chi_{g}^e$}, which is the number of CZ time steps needed to prepare that entanglement class ($r(\chi_{i}, \; \chi_{g}^e) = \{0.032\pm 0.04, -0.09 \pm 0.09\}$).

Meanwhile, orbit chromatic index, \ch{$\chi_{i}^e$}, and \ch{Schmidt measure}, \ch{$E_S$}, correlate differently, depending on whether isomorphic graph states are considered equal ($r(\chi_{i}^e, \; E_S) = \{0.81 \pm 0.04, 0.44 \pm 0.11\}$ for $n\leq8$ and $n\leq7$ respectively).
Chromatic index indicates gives a lower bound on the maximum degree in the orbit, that is, the maximum number of graph states that a single graph of the orbit is at most $\chi_i^e$.
Hence classes with high Schmidt measure tend to have at least one graph state that can produce many different graphs states by local complementation.
In most cases, $\chi_{L_i}^e=n$ because the $n$ available local complementations produce different (but potentially isomorphic graphs).
Hence little information can be gained from the chromatic index of $L_i$ orbits, $\chi_{L_i}^e$.\makeatletter\onecolumngrid@push\makeatother
\begin{table*}[p]
\centering
\small{
\begin{adjustbox}{center}
\begin{tabular}{P{0.7cm}	P{0.4cm}		P{0.34cm}	P{0.8cm}		P{0.55cm}		P{0.45cm}		P{0.55cm}			P{1.2cm}		P{0.35cm}		P{0.35cm}			P{0.35cm}			P{0.35cm}		P{0.72cm}		P{0.7cm}						P{1.15cm}					 P{0.68cm}				P{0.44cm}		P{0.64cm}	P{0.3cm}	P{0.3cm}}

Class					& $|Q|$			&	$|e|$	&	$E_S$ 	&$\mathrm{rwd}$ 	&	$|C_i|$ 	&	$|E_i|$ 	&	$|E_i|/|C_i|$ 	&	$\chi_g$	&	$\chi_g^e$		&	$\chi_{C_i}$	&	$\chi_{C_i}^e$	&	Tree		&	$\langle d_{jk}^{C_i}\rangle$ 	&$\mathrm{max}(d_{jk}^{C_i})$	&$|\mathrm{aut}|$	& 	2D			&	Loop 	&	E.		&	H.
\\
\noalign{\vskip 2mm}    
\hline
\hline
\noalign{\vskip 2mm}
\hspace{0mm}3			&		4		&	3		&	1		&	1				&		2		&		2		&		1			&		2		&	3				&		2			&		1		&	\cmark		&		1							&		1						&		1			&	\cmark		&	\cmark	&	\xmark	&	\xmark\\ 
\hspace{0mm}4			&		4		&	3		&	2		&	1				&		4		&		5		&		1.25		&		2		&	2				&		2			&		2		&	\cmark		&		1.67						&		3						&		1			&	\cmark		&	\cmark	&	\xmark	&	\xmark\\ 
\noalign{\vskip 2mm}    

\noalign{\vskip 2mm}  
\hspace{0mm}5			&		5		&	4		&	1		&	1				&		2		&		2		&		1			&		2		&	4				&		2			&		1		&	\cmark		&		1							&		1						&		1			&	\cmark		&	\cmark	&	\xmark	&	\xmark\\ 
\hspace{0mm}6			&		5		&	4		&	2		&	1				&		6		&		9		&		1.5			&		2		&	3				&		2			&		2		&	\xmark		&		1.8							&		3						&		2			&	\cmark		&	\cmark	&	\cmark	&	\cmark\\ 
\hspace{0mm}7			&		5		&	4		&	2		&	1				&		10		&		19		&		1.9			&		2		&	2				&		3			&		3		&	\xmark		&		2.04						&		3						&		1			&	\cmark		&	\cmark	&	\xmark	&	\cmark\\ 
\hspace{0mm}8			&		5		&	5		&	$2<3$	&	2				&		3		&		3		&		1			&		3		&	3				&		2			&		2		&	\cmark		&		1.33						&		2						&		1			&	\cmark		&	\cmark	&	\xmark	&	\xmark\\ 
\noalign{\vskip 2mm}   

\noalign{\vskip 2mm}  
\hspace{0mm}9			&		6		&	5		&	1		&	1				&		2		&		2		&		1			&		2		&	5				&		2			&		1		&	\cmark		&		1							&		1						&		1			&	\cmark		&	\cmark	&	\xmark	&	\xmark\\ 
\hspace{0mm}10			&		6		&	5		&	2		&	1				&		6		&		9		&		1.5			&		2		&	4				&		2			&		2		&	\xmark		&		1.8							&		3						&		2			&	\cmark		&	\cmark	&	\cmark	&	\cmark\\ 
\hspace{0mm}11			&		6		&	5		&	2		&	1				&		4		&		5		&		1.25		&		2		&	3				&		2			&		2		&	\cmark		&		1.67						&		3						&		1			&	\cmark		&	\cmark	&	\xmark	&	\xmark\\ 
\hspace{0mm}12			&		6		&	5		&	2		&	1				&		16		&		34		&		2.13		&		2		&	3				&		3			&		3		&	\xmark		&		2.25						&		3						&		3			&	\xmark		&	\cmark	&	\xmark	&	\cmark\\ 
\hspace{0mm}13			&		6		&	5		&	3		&	1				&		10		&		20		&		2			&		2		&	3				&		3			&		3		&	\xmark		&		2.04						&		3						&		1			&	\cmark		&	\cmark	&	\xmark	&	\cmark\\ 
\hspace{0mm}14			&		6		&	5		&	3		&	1				&		25		&		58		&		2.32		&		2		&	2				&		3			&		4		&	\xmark		&		2.51						&		5						&		2			&	\xmark		&	\cmark	&	\cmark	&	\xmark\\ 
\hspace{0mm}15			&		6		&	6		&	2		&	1				&		5		&		8		&		1.6			&		2		&	3				&		3			&		3		&	\xmark		&		1.7							&		3						&		1			&	\cmark		&	\cmark	&	\xmark	&	\xmark\\ 
\hspace{0mm}16			&		6		&	6		&	3		&	1				&		5		&		9		&		1.8			&		3		&	3				&		3			&		3		&	\xmark		&		1.7							&		3						&		1			&	\cmark		&	\cmark	&	\xmark	&	\xmark\\ 
\hspace{0mm}17			&		6		&	6		&	3		&	2				&		21		&		47		&		2.24		&		3		&	3				&		3			&		5		&	\xmark		&		2.32						&		4						&		0			&	\xmark		&	\cmark	&	\xmark	&	\cmark\\ 
\hspace{0mm}18			&		6		&	6		&	3		&	2				&		16		&		29		&		1.81		&		2		&	2				&		3			&		6		&	\xmark		&		2.22						&		4						&		0			&	\xmark		&	\cmark	&	\xmark	&	\xmark\\ 
\hspace{0mm}19			&		6		&	9		&	$3<4$	&	2				&		2		&		2		&		1			&		3		&	3				&		2			&		1		&	\cmark		&		1							&		1						&		1			&	\cmark		&	\cmark	&	\xmark	&	\xmark\\ 
\noalign{\vskip 2mm} 

\noalign{\vskip 2mm}  
\hspace{0mm}20			&		7		&	6		&	1		&	1				&		2		&		2		&		1			&		2		&	6				&		2			&		1		&	\cmark		&		1							&		1						&		1			&	\cmark		&	\cmark	&	\xmark	&	\xmark\\ 
\hspace{0mm}21			&		7		&	6		&	2		&	1				&		6		&		9		&		1.5			&		2		&	5				&		2			&		2		&	\xmark		&		1.8							&		3						&		2			&	\cmark		&	\cmark	&	\cmark	&	\cmark\\ 
\hspace{0mm}22			&		7		&	6		&	2		&	1				&		6		&		9		&		1.5			&		2		&	4				&		2			&		2		&	\xmark		&		1.8							&		3						&		2			&	\cmark		&	\cmark	&	\cmark	&	\cmark\\ 
\hspace{0mm}23			&		7		&	6		&	2		&	1				&		16		&		34		&		2.13		&		2		&	4				&		3			&		3		&	\xmark		&		2.25						&		3						&		3			&	\xmark		&	\cmark	&	\xmark	&	\cmark\\ 
\hspace{0mm}24			&		7		&	6		&	2		&	1				&		10		&		19		&		1.9			&		2		&	3				&		3			&		3		&	\xmark		&		2.04						&		3						&		1			&	\cmark		&	\cmark	&	\xmark	&	\cmark\\ 
\hspace{0mm}25			&		7		&	6		&	3		&	1				&		10		&		20		&		2			&		2		&	4				&		3			&		3		&	\xmark		&		2.04						&		3						&		1			&	\cmark		&	\cmark	&	\xmark	&	\cmark\\ 
\hspace{0mm}26			&		7		&	6		&	3		&	1				&		16		&		35		&		2.19		&		2		&	3				&		3			&		3		&	\xmark		&		2.25						&		3						&		3			&	\xmark		&	\cmark	&	\xmark	&	\cmark\\ 
\hspace{0mm}27			&		7		&	6		&	3		&	1				&		44		&		114		&		2.59		&		2		&	3				&		3			&		4		&	\xmark		&		2.84						&		5						&		3			&	\xmark		&	\cmark	&	\cmark	&	\cmark\\ 
\hspace{0mm}28			&		7		&	6		&	3		&	1				&		44		&		118		&		2.68		&		2		&	3				&		3			&		4		&	\xmark		&		2.84						&		5						&		3			&	\xmark		&	\cmark	&	\cmark	&	\cmark\\ 
\hspace{0mm}29			&		7		&	6		&	3		&	1				&		14		&		30		&		2.14		&		2		&	3				&		3			&		4		&	\xmark		&		2.34						&		5						&		1			&	\cmark		&	\cmark	&	\xmark	&	\cmark\\ 
\hspace{0mm}30			&		7		&	6		&	3		&	1				&		66		&		191		&		2.89		&		2		&	2				&		3			&		5		&	\xmark		&		3.05						&		6						&		2			&	\xmark		&	\cmark	&	\xmark	&	\cmark\\ 
\hspace{0mm}31			&		7		&	7		&	2		&	1				&		10		&		20		&		2			&		2		&	4				&		3			&		3		&	\xmark		&		2.04						&		3						&		1			&	\cmark		&	\cmark	&	\xmark	&	\cmark\\ 
\hspace{0mm}32			&		7		&	7		&	3		&	1				&		10		&		21		&		2.1			&		3		&	4				&		3			&		3		&	\xmark		&		2.04						&		3						&		1			&	\cmark		&	\cmark	&	\xmark	&	\cmark\\ 
\hspace{0mm}33			&		7		&	7		&	3		&	2				&		21		&		47		&		2.24		&		3		&	4				&		3			&		5		&	\xmark		&		2.31						&		4						&		0			&	\xmark		&	\cmark	&	\xmark	&	\cmark\\ 
\hspace{0mm}34			&		7		&	7		&	3		&	1				&		26		&		68		&		2.62		&		2		&	3				&		3			&		4		&	\xmark		&		2.50						&		4						&		3			&	\xmark		&	\cmark	&	\xmark	&	\cmark\\ 
\hspace{0mm}35			&		7		&	7		&	3		&	2				&		36		&		98		&		2.72		&		3		&	3				&		3			&		5		&	\xmark		&		2.54						&		4						&		1			&	\xmark		&	\cmark	&	\xmark	&	\cmark\\ 
\hspace{0mm}36			&		7		&	7		&	3		&	1				&		28		&		70		&		2.5			&		3		&	3				&		3			&		4		&	\xmark		&		2.62						&		5						&		3			&	\xmark		&	\cmark	&	\xmark	&	\cmark\\ 
\hspace{0mm}37			&		7		&	7		&	3		&	2				&		72		&		206		&		2.86		&		3		&	3				&		3			&		5		&	\xmark		&		3.06						&		5						&		2			&	\xmark		&	\cmark	&	\xmark	&	\cmark\\ 
\hspace{0mm}38			&		7		&	7		&	3		&	2				&		114		&		336		&		2.94		&		2		&	3				&		3			&		6		&	\xmark		&		3.29						&		6						&		2			&	\xmark		&	\cmark	&	\cmark	&	\cmark\\ 
\hspace{0mm}39			&		7		&	7		&	$3<4$	&	2				&		56		&		157		&		2.80		&		3		&	3				&		4			&		6		&	\xmark		&		2.85						&		5						&		1			&	\xmark		&	\cmark	&	\xmark	&	\cmark\\ 
\hspace{0mm}40			&		7		&	7		&	$3<4$	&	2				&		92		&		271		&		2.95		&		3		&	3				&		3			&		7		&	\xmark		&		3.02						&		7						&		1			&	\xmark		&	\xmark	&	\xmark	&	\xmark\\ 
\hspace{0mm}41			&		7		&	8		&	$3<4$	&	2				&		57		&		164		&		2.88		&		3		&	3				&		3			&		6		&	\xmark		&		2.79						&		5						&		1			&	\xmark		&	\cmark	&	\xmark	&	\cmark\\ 
\hspace{0mm}42			&		7		&	8		&	$3<4$	&	2				&		33		&		80		&		2.42		&		3		&	3				&		5			&		7		&	\xmark		&		2.43						&		5						&		0			&	\xmark		&	\cmark	&	\xmark	&	\cmark\\ 
\hspace{0mm}43			&		7		&	9		&	3		&	2				&		9		&		16		&		1.78		&		2		&	3				&		3			&		5		&	\xmark		&		1.81						&		3						&		1			&	\cmark		&	\cmark	&	\xmark	&	\xmark\\ 
\hspace{0mm}44			&		7		&	9		&	$3<4$	&	2				&		46		&		109		&		2.37		&		3		&	3				&		5			&		7		&	\xmark		&		2.81						&		5						&		0			&	\xmark		&	\cmark	&	\xmark	&	\cmark\\ 
\hspace{0mm}45			&		7		&	10		&	$3<4$	&	2				&		9		&		16		&		1.78		&		3		&	4				&		3			&		4		&	\xmark		&		1.97						&		4						&		0			&	\cmark		&	\cmark	&	\xmark	&	\xmark\\

\noalign{\vskip 2mm}   
\end{tabular}
\end{adjustbox}
}

\caption{\lineskip=0pt A selection of properties of $C_i$ (see Appendix Figure \ref{img:bigtable} for a table showing a representative graph state from each class of $n<9$ qubits). Here, $|Q|$ is the number of qubits of the orbit's graph states, $|e|$ is smallest number of edges of any graph state member of $C_i$. Each class's \ch{Schmidt measure}, $E_S$, is written $a < b $ to compactly express lower ($a$) and upper ($b$) bounds, when an exact value is not known\cite{hein2004multiparty, cabello2011optimal}. $\mathrm{rwd}$ is the class's rank-width, $|C_i|$ is the size of the orbit, $|E_i|$ is the number of edges on the orbit, $\chi_g$ is the minimum chromatic number of the graph states in the class,  $\chi_g^e$ is the minimum chromatic index of the graph states in the class (which corresponds to the minimum number of CZ gates required to prepare them), $\chi_{C_i}$ is the orbits chromatic number, $\chi_{C_i}^e$ is the orbits chromatic index, `Tree' is whether the orbit is a tree (excluding self-loops), \ch{$d_{jk}$ are the distances between vertices of the orbit (therefore $\langle d_{jk}^{C_i} \rangle$ is the mean distance between any two vertices and $\mathrm{max}(d_{jk}^{C_i})$ the diameter of the orbit), $|\mathrm{aut}|$ is the size of the automorphism group of the orbit}, `2D' is whether the orbit is planar, `Loop' is whether the orbit has any self-loops, `E.' (`H.') is whether the graph has a cycle in which each edge (vertex) of the orbit is visited precisely once. Definitions of all of these quantities can be found in Appendix Section \ref{sec:defs}. Properties of $L_i$ orbits may differ to their $C_i$ partner. \medskip
}
\label{tab:orbittab}

\end{table*}

\clearpage
\makeatletter\onecolumngrid@pop\makeatother\noindent$\chi_{C_i}^e$, however, is much more varied and correlates well with Schmidt rank. 
This can be understood as $C_i$ orbits consider only the topological properties of the graph states.

We note that \ch{Schmidt measure}, \ch{$E_S$} correlates well with rank-width, \ch{$\mathrm{rwd}$}, and minimum edge count, \ch{$|e|$}, ($r(E_S, \; \mathrm{rwd})= 0.62\pm0.03$, $r(E_S, \; |e|) = 0.78 \pm 0.02$), but not with graph state chromatic index, \ch{$\chi_{g}^e$}:  ($r(E_S, \; \chi_{g}^e) = -0.17 \pm 0.02$).
Interestingly, Schmidt \ch{measure}, \ch{$E_S$}, (and therefore orbit chromatic index, \ch{$\chi_{i}^e$}), \ch{strongly correlates with minimum edge count, \ch{$|e|$}, (the total number of CZs required to prepare an entanglement class) but not with graph state chromatic index, \ch{$\chi_{g}^e$}, (the number of CZ time steps required to prepare an entanglement class)}. 
\ch{Resources for quantum computation are often lattices, and hence have constant CZ preparation complexity (in terms of time steps), though their rank-width must grow faster than logarithmically\cite{van2007classical}.}
\ch{We also note that there exist efficient entanglement purification protocols for graph states which have a chromatic number $\chi_{g}=2$ (those which are two-colourable)\cite{dur2003multiparticle}.}
\ch{We note that all of the correlations we observe with $r>\nicefrac{1}{2}$ appear to come from to be well behaved distributions, with no `\jer{catestrophic} failures' observed.}

Some properties of a entanglement class' graph states can be deduced from properties of their orbit.
For example, class no.~$40$ has is the only orbit up to seven qubits which has no self-loops.
This implies that none of its member graph states have a vertex of degree $1$ (leaves).
Increasing qubit number, we observe that $9$\% of orbits with $n\leq10$ qubits do not have self-loops.
\ch{It follows that the number of CZ gates needed to generate any member of these classes is at least $n$.}

Local complementation commutes when the neighbourhoods of the two indices are disjoint.
This creates a cycle in the graph state's orbit.
\ch{Hence it can be deduced that orbits which are trees only contain graph states in which all vertices are at most distance two from one another, since they must share part of their neighbourhood with every other qubit.}
We note that only Greenberger-Horne-Zeilinger (GHZ) entanglement gives rise to $L_i$ orbits that are trees (for $n\leq 8$), and these contain only two \ch{non-isomorphic} graph states.
Meanwhile there is one 3-vertex orbit and three 4-vertex $C_i$ orbits which are trees for $n\leq10$ qubits.
These are connected in a line, and contain self-loops (see Fig.~\ref{img:line4ghz4}e).

\begin{figure*}[t]
\centering

\captionsetup{width=1.0\textwidth}
\includegraphics[width=1.0\textwidth,center]{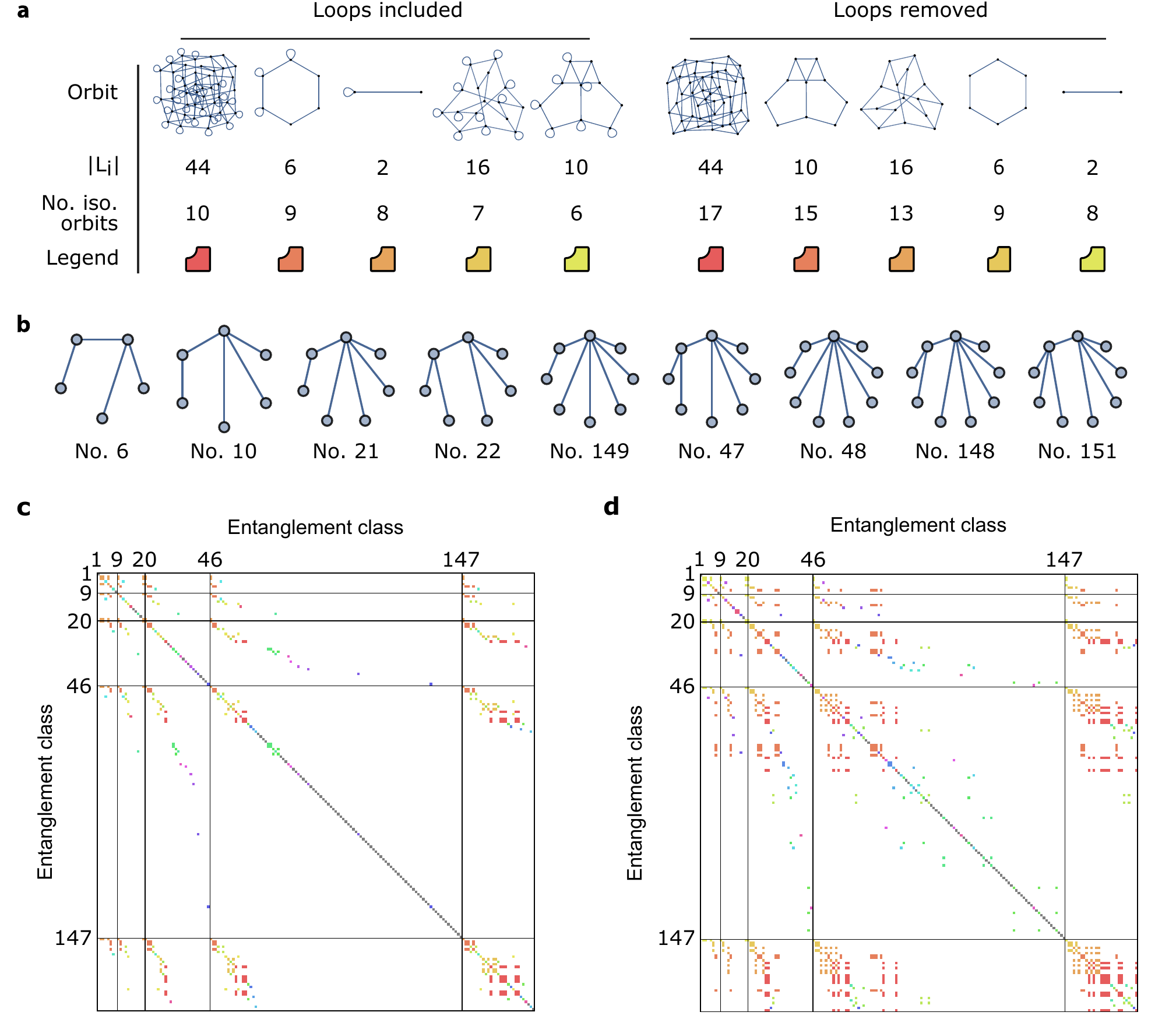}

\caption{Isomorphism of local complementation orbits. \textbf{a.} The five most common orbits up to class $150$, considering orbit self-loops and not. \textbf{b.} Canonical, minimum-edge representatives of the orbits $C_i$ for $i = 6, 10, 21, 22, 47, 48, 148, 149, 151$, each of which are isomorphic to one another (an order-six ring with three adjacent self-loops, see \textbf{a}). \textbf{c.} Isomorphism of orbits $C_i$. $I_{ij} = 1$ if orbit $C_i$ and $C_j$ are isomorphic. Entries are coloured by isomorphism.  Regions of equal qubit number are demarcated. \textbf{d.} Isomorphism of $C_i$ orbits with self-loops removed.
}
\label{img:iso}
\end{figure*}

Interestingly, some $C_i$ orbits are isomorphic to other orbits, $C_j$ ($i\neq j$).
Fig.~\ref{img:iso}a shows a table of the most commonly found orbits, their size, and their frequency in our dataset.
Furthermore, Fig.~\ref{img:iso}c displays which orbits are isomorphic to one another in the form of a matrix.

\ch{Figs.~\ref{img:line4ghz4}c and \ref{img:line4ghz4}f are examples of isomorphic $C_i$ orbits with vastly different entanglement properties, that is, perfect correlation (GHZ entanglement) and an optimal error correcting code\cite{schlingemann2001quantum}.
In contrast, there are no $L_i$ orbits that are isomorphic for $n\leq8$.
That is, only if the graph states are local Clifford equivalent are their $L_i$ orbits isomorphic. 
It is unclear which properties of a graph state lead to isomorphic orbits, however we note that graphs which share an isomorphic orbit often---but not always---have a similar connectivity.
Fig.~\ref{img:iso}b displays a set of similar but distinct graph states whose orbits are isomorphic}

\ch{A simple example of isomorphic orbits comes from $n$-qubit GHZ entanglement, which always contain the $n$-qubit `star' graph.
%
%
In a star graph, there are only two different local complementation operations.
That is, local complemenation can be applied to the centre qubit or to one of the leaves. 
Applying local complementation to the leaves does nothing, while applying local complementation to the centre qubit yields the fully connected graph state. 
Applying local complementation to qubit $\alpha$ of the fully-connected state yields a star graph state where the center of the star is qubit $\alpha$.
Hence these orbits contain only the star and the fully-connected graph states.
$L_i$ orbits have all $n$ of the isomorhpisms of the star graph state connected to the fully-connected graph state, and are themselves in an $n+1$ vertex star formation (see Fig.~\ref{img:line4ghz4}b), while $C_i$ orbits have only two members for all $n$ (see Fig.~\ref{img:line4ghz4}c)}.

The proportion of all graphs which are asymmetric tends towards zero as the number of vertices tends towards infinity\cite{erdHos1963asymmetric}, (${\sim}50\%$ of unlabelled 9-vertex graphs)
However, the majority of the orbits we compute are symmetric ($75$\% of 9-qubit orbits have a non-empty automorphism group), including orbits containing thousands of graph states.
\ch{A study of the symmetries of the orbits, which is quantified by the size of the orbit's automorphism group, $|\mathrm{aut}|$, is left for future research.}

Many of the computed parameters, such as \ch{Schmidt measure}, rank-width and automorphism group have exponential complexity with system size.
The rank-width, while exponential in nature, can be computed exactly\cite{oum2009computing}, while the \ch{Schmidt measure} requires a nonconvex, nonlinear optimisation, and so is more challenging.
We rely on previously computed\cite{cabello2011optimal, danielsen2006classification} bounds of the \ch{Schmidt measure}, while computing the rank-width using the software `SAGE'\cite{sage}.
Though our graph states are small, there is an exponential number of entanglement classes as qubit number is increased.
Further, many of the graph metrics discussed, such as graph colouring (chromatic number and chromatic index) belong to complexity class \textsc{NP}.
As such, they become challenging to compute on dense orbits with thousands of vertices.
For this reason we computed the chromatic index only for $n\leq8$ and $n\leq7$ for $C_i$ and $L_i$ orbits respectively.
All graph colouring computations were performed with the software `IGraph/M'\cite{igraph, igraphm}.

Due to their connectivity and scale, the majority of orbits we explored are far too complex to view directly, as we did in Fig.~\ref{img:line4ghz4}.
We can instead represent them with matrices.
Fig.~\ref{img:adjmat-6-2} shows the adjacency matrices and distance matrices of class $L_{10}$.
We order the matrix by isomorphism, edge count, and then lexicographically by their lexicographically sorted edgelists.
Further, we demarcate regions of the plot which correspond to graph states that have the same number of edges and that are isomorphic to one another for $C_i$ and $L_i$ respectively.
In both cases, the adjacency matrices show structure related to these regions.

There is variety and scale in the 587 $C_i$ orbits and 147 $L_i$ orbits we have computed which cannot be reproduced in a single article.
A curated selection of orbits is displayed in Appendix Section \ref{sec:gallery}, and the full data set is available online.

\section{Discussion}

It is likely that future quantum information processors will have restricted two-qubit gate topology, due to the qubit's physical locations and proximity.
Since single-qubit operations are commonly faster or higher-fidelity than two-qubit gates, local complementation may be used to improve a device's speed or fidelity\cite{adcock2018hard}.
For a prescriptive method, the relationship between orbits by nonlocal CZ gates must be known.
A complete map of this type would describe how all $n$-qubit graph states are related to one another, and provide a look up table for optimal transformations between them.
From here, the addition of vertex deletion would give a complete map of graph states under LC+LPM+CC operations (the vertex minor problem).
A doubly-exponential problem, computation of these maps appears to be infeasible for even modest $n$.
For small graphs, however, such a map may be enlightening---the exploration is left for future work.

Knowledge of the orbits of local complementation may also enable in quantum secret sharing and quantum networks\cite{markham2008graph, hahn2018quantum}.
A graph state may be distributed between separated parties, each of whom can perform local operations and communicate with their neighbours (according to the graph state structure).
This allows different quantum protocols to be implemented using a resource which has already been distributed spatially.
If the parties only have knowledge of their own neighbourhood, and each party performs local complementation at random, the shared state can be scrambled.
Numerically, we find the stationary distributions generated by random walks on the orbits appear to tend towards uniform as orbit size increases, implying this `scrambling' is effective.
This could be formalised further by investigating mixing rates.

Local complementation allows the entanglement of a resource state to be utilised differently in measurement-based protocols\cite{joo2011edge, zwerger2012measurement, hahn2018quantum}.
That is, a resource state can be transformed into any other state from its entanglement class, and used according to its shape.
Though practically this simply corresponds to changing the protocol measurement bases, considering LC-equivalent graph states as a new state preserves the standard language of measurement-based protocols (measurement in the $X$-$Y$ plane and $Z$ directions).
Generally, local complementation has merit in applications where qubits are in inequivalent spatial locations---it  illustrates the many functions of a given entanglement.

In some quantum computer architectures, such as those for linear optical quantum computing\cite{gimeno2015three}, percolated resource states are generated probabilistically.
These states have a randomly generated structure, and hence some are more powerful than others, for example they may have more favourable connectivity for pathfinding\cite{morley2017physical} or loss tolerance\cite{rudolph2017optimistic, morley2018loss}, which may be optimised by local complementation.
Though the entanglement class of any useful resource state will be too large to compute directly, it may be possible to develop heuristics for using local complementation to optimise local regions of the resource.
These heuristics may be explored and verified with the algorithm of ref.~\citenum{van2004efficient}.
\ch{In this sense, optimisation via local complementation can be seen as a step in the compilation of a protocol or algorithm given a specific hardware.}

Our library of orbits, including the code used to generated the plots in this manuscript (Mathematica) is available online\cite{adcock2019graphorbitsonline} and comprises 35 MB compressed.
\ch{We also provide\cite{gsc} a new software tool, `graph state compass', which computes the orbit of any input graph state (python).}
Exploration up to $n=12$, where representative graph states of each orbit are known, is feasible if a compiled language and parallelism are employed. 
Extending the database further is a significant computational challenge, as, though an exact scaling is not known, the number of graph state entanglement classes grows super-exponentially for $n\leq 12$ qubits

Our exploration opens new lines of enquiry in the study of graph states and their entanglement.
\ch{For example, what can be learned from the symmetries of an orbit?}
Is it possible to completely map LC+LPM+CC operations beyond 12 qubits?
What new applications are possible utilising knowledge of LC orbits?

Stabiliser state entanglement is---and will continue to be---at the core of quantum information protocols. 
The resource we provide gives a new handle to investigate the rich relationship between graph theory, stabiliser state entanglement, and applications of quantum information.


\section*{Acknowledgements}

We would like to thank Caterina Vigliar, Sam Pallister, Will McCutcheon, and John G. Rarity for their invaluable help. \ch{We would also like to thank the reviewers of this paper for their thorough examination of our work and for giving us inspiration to improve it}. The computer programs `IGraph/M' and `SAGE' were vital for evaluating difficult to compute properties of our extensive library of graphs. This work was supported by EPSRC Programme Grant EP/L024020/1, the EPSRC Quantum Engineering Centre for Doctoral Training EP/L015730/1, and the ERC Starting Grant ERC-2014-STG 640079. JWS acknowledges the generous support of the Leverhulme Trust, through Leverhulme Early Career Fellowship ECF-2018-276.


\setcounter{figure}{0}
\renewcommand{\figurename}{Appendix Figure}

\cleardoublepage

\setcounter{section}{0}

\onecolumn

\section*{Appendix}
\label{sec:appen}


\ch{
\section{Definition of quantities}
\label{sec:defs}
In this section, we introduce and define the graph-theoretical quantites and concepts used in this paper. Firstly, a graph, $G(V,E)$, is composed of a set of vertices $V$, and a set of edges $E$. Edges are two-element subsets of $V$. In directed graphs, the elements of $E$ are ordered. An example graph is the (undirected) fully-connected three qubit graph, for which $V=\{1,2,3\}$ and $E=\{(1,2),(2,3),(1,3)\}$. The following is a glossary of terms and definitions of properties of a graph $G(V,E)$:

\begin{itemize}

\item\emph{Adjacent}. Two vertices $v_1$ and $v_2$ are adjacent if there is a edge $e$ which contains $v_1$ and $v_2$ as elements. 

\item \emph{Walk}. A walk is an alternating sequence of edges and vertices, e.g.~ $W=v_1e_{12},v_2\ldots v_j$

\item \emph{Path}. A path is a walk where no vertex is revisited

\item \emph{Cycle}. A cycle is a walk in which has one repeated vertex, which is the the first and last vertices of the walk.

\item \emph{Tree}. A tree is a graph which has no cycles. Trees necesarily have bounded rank-width, and can not be resources for universal quantum computation\cite{van2007classical}.

\item \emph{Neighbourhood}. The neighbourhood of a vertex, $\alpha$, of a graph, G, is written $N_{G}(\alpha)$. This is the set of vertices with which the vertex $\alpha$ is adjacent to.

\item \emph{Leaf}. A leaf is a vertex of a graph which is adjacent to precisely one other vertex. Leaves can be removed from graph states using a $Z$ measurement without disturbing the rest of the graph state.

\item \emph{self-loop}. A self-loop is an edge from a vertex to itself: $e=(v_i,v_i)$.

\item \emph{Connected}. A graph is connected if there is a path from every vertex to every other vertex. Connected graph states are globally entangled (have genuine multipartite entanglement). Every orbit induced by local complementation is connected.

\item \emph{Connected}. A graph, $G(V,E)$ is connected if every vertex of the graph is adjacent to every other vertex: $\forall i,j \in V, \; (i,j)\in E$. Fully-connected graph states have GHZ-type entanglement.

\item \emph{Subgraph}. The graph $G'=(V',E')$ is a subgraph of $G(V,E)$ if $V'\subset V $ and $\forall e' \in E', e'\in E$. For a graph $G(V,E)$, the neighbourhood of a vertex $\alpha$, $N_{G}(\alpha)$ is a subgraph of $G$

\item \emph{Adjacency matrix}. The adjacency matrix, $A$, of a graph, $G$, has elements $a_{ij} = 1$ if $(i,j)\in E$. Otherwise $a_{ij} = 0$.

\item \emph{Distance}. The distance, $d^G_{jk}$ between vertices $i$ and $j$ of a graph $G(V,E)$ is the number of edges in the shortest path between vertex $i$ and vertex $j$. In an orbit, the distance between two graph states $\ket{\psi_1}$ and $\ket{\psi_2}$ is the number of local complementations needed to transform $\ket{\psi_1}$ into $\ket{\psi_2}$.

\item \emph{Distance matrix}. The distance matrix, $D$, of a graph, $G$, has elements $d_{jk} = d^G_{jk}$ and $d^G_{jj}=0$. Otherwise $d^G_{jk} = 0$.

\item \emph{Diameter}. The diameter of a graph $G(V,E)$ is the largest distance $d^G_{jk}$ of that graph. This is the maximum number of local complementations needed to transform one graph state into another within an orbit.

\item \emph{Chromatic index}. The chromatic index is the minimum number of colours needed to colour the edges of a graph, so that no two edges of the same colour are adjacent. This corresponds to the preparation complexity of a graph state using only CZs in terms of number of time steps.

\item \emph{Chromatic number}. The chromatic number is the minimum number of colours needed to colour the vertices of a graph, so that no two vertices of the same colour are adjacent. This corresponds to the preparation complexity of a graph state using only CZs in terms of number of number of CZ gates.

\item \emph{Planar graph}. A graph is planar if it can be drawn on a two-dimensional plane without any of its edges crossing.

\item \emph{Eulerian cycle}. A Eulerian cycle is a cycle on a graph which traverses every edge of that graph precisely once.

\item \emph{Hamiltonian cycle}. A Hamiltonain cycle is a cycle on a graph which visits every vertex of that graph precisely once.

\item \emph{Automorphism group}. The automorphism group of a graph is the set of vertex permutations (relabellings) that transform a graph into itself. This is the set of symmetries of the graph.

\item \emph{Cut-rank}\cite{dahlberg2018transforming}. For some $A \subset V$, the cut-rank of an adjacency matrix $\Gamma$ with respect to $A$ is:
\begin{equation}
    \mathrm{cutrk}(\Gamma)  \vcentcolon = \mathrm{rank}_{\mathbb{F}_2}(\Gamma[A,V \setminus A]),
\end{equation}
where $\Gamma[A,V \setminus A]$ is the submatrix of $\Gamma$ obtained by taking rows $A$ and columns $V \setminus A$. Here, the rank is taken over the finite field of order 2 (addition modulo 2). Note that the cut-rank with respect to $A$ of a graph $G$ is equal to the Schmidt-rank of
the state $G$ with respect to the bipartition $(A, V\setminus A)$\cite{dahlberg2018transforming}. This is used in the definition of rank decomposition, and finally rank-width below.

\item \emph{Rank decomposition}\cite{dahlberg2018transforming}. A rank decomposition is a pair $R = (T, \mu)$ where $T$ is a subcubic tree and $\mu$ is a bijection $\mu : V(g) \rightarrow {l : l \;\mathrm{is \; a \; leaf \; of} \; T}$. A subcubic tree is a tree with at least two vertices where each vertex has degree less than or equal to 3. Deleting any edge $e$ in $T$ splits the tree into two connected components and therefore induces a partition $(A_e, \;  B_e)$ of the leaves. The width of an edge $e$ of the subcubic tree is defined as the cut-rank of the corresponding partition. Furthermore, the width of the rank-decomposition is defined as the maximum width over all edges, i.e:
\begin{equation}
    \mathrm{width}_R \vcentcolon = \max_{e \in E(T)} \mathrm{cutrk}_{\mu^{-1}(T,\;e)}(G).
\end{equation}
Which is used in the definition of rank-width below.

\item \emph{rank-width}\cite{dahlberg2018transforming}. Utilising the above definitions, the rank-width is as follows:
\begin{equation}
    \mathrm{rwd} \vcentcolon = \min_{R} \mathrm{width}_R(G) = \min_{R}(G)\min_{T,\;\mu}\max_{e\in E(T)}\mathrm{cutrk}_{\mu^{-1}(T,e)}(G).
\end{equation}
Any graph state property which is expressible in so-called monadic second-order logic (a higher-order logical system) can be computed in time $O(f(\mathrm{rwd}(G))|V(G)|^3)$, where $f$ is an exponential function\cite{dahlberg2019complexity}. 
This includes the vertex minors problem, (deciding whether a graph can be transformed into another via LC operations, local pauli measurement and classical communication).

\end{itemize}
\vspace{0.5cm}
\noindent Below, We reproduce the definitions of some of the quantum information quantities discussed in the main text:

\begin{itemize}
\item \emph{\ch{Schmidt measure}}\cite{eisert2001schmidt}. The minimum number of terms in which a quantum state can be represented by, over all local bases. We can write an $n$-partite system with parties $A_i,\ldots,A_n$, each with a quantum system of dimension $d_{i},\ldots,d_n$ as the following:
\begin{equation}
    \ket{\psi} = \sum_i^R\alpha_i\ket{\psi^{(1)}_{A_i}} \otimes \ldots \otimes \ket{\psi^{(n)}_{A_n}}.
\end{equation}
Where $\ket{\psi^{(j)}_{A_i}}$ are local states in the $i^\mathrm{th}$ subspace of $i=1\ldots R$. Let $r$ be the minimum possible $R$ across all possible choices of local basis $\ket{\psi^{(j)}_{A_i}}$ for each of the subspaces $A_i$ (minimising basis choice $j$). The \ch{Schmidt measure} is then $\mathrm{log}_2(r)$.

\item \emph{Schmidt-rank-width}\cite{dahlberg2018transforming, van2007classical}. Starting with the graph state $\ket{\psi}$, we define Schmidt-rank-width. Letting $\chi_{(A^e_T,\; B^e_T)}$ be the number of nonzero Schmidt coefficients of $\ket{\psi}$ with respect to the bipartition $(A^e_T,B^e_T)$ of $V$, the qubits of $\ket{\psi}$
\begin{equation}
    \chi_\mathrm{wd} \vcentcolon = \min_{T,\mu}\max_{e\in E(T)} \mathrm{log}_2( \chi_{A^e_TB^e_T}(\ket{\psi}))
\end{equation}
For a subcubic tree $T$ with edge $e$ around which a bipartition is formed (as in the definition of rank-width). The Schmidt-rank-width of the graph state $\ket{G}$ is equal to the rank-width of a graph $G$. 

\end{itemize}

}

\newpage
\section{A gallery of graph state orbits}
\label{sec:gallery}

We map over 600 orbits, most of which are vastly too large and complex to display. In this Appendix, we exhibit a curated selection of orbits to demonstrate the variety of form they display.
\vspace{5cm}

\begin{figure*}[h]
\centering
\captionsetup{width=1.0\textwidth}
\includegraphics[width=1.0\textwidth,center]{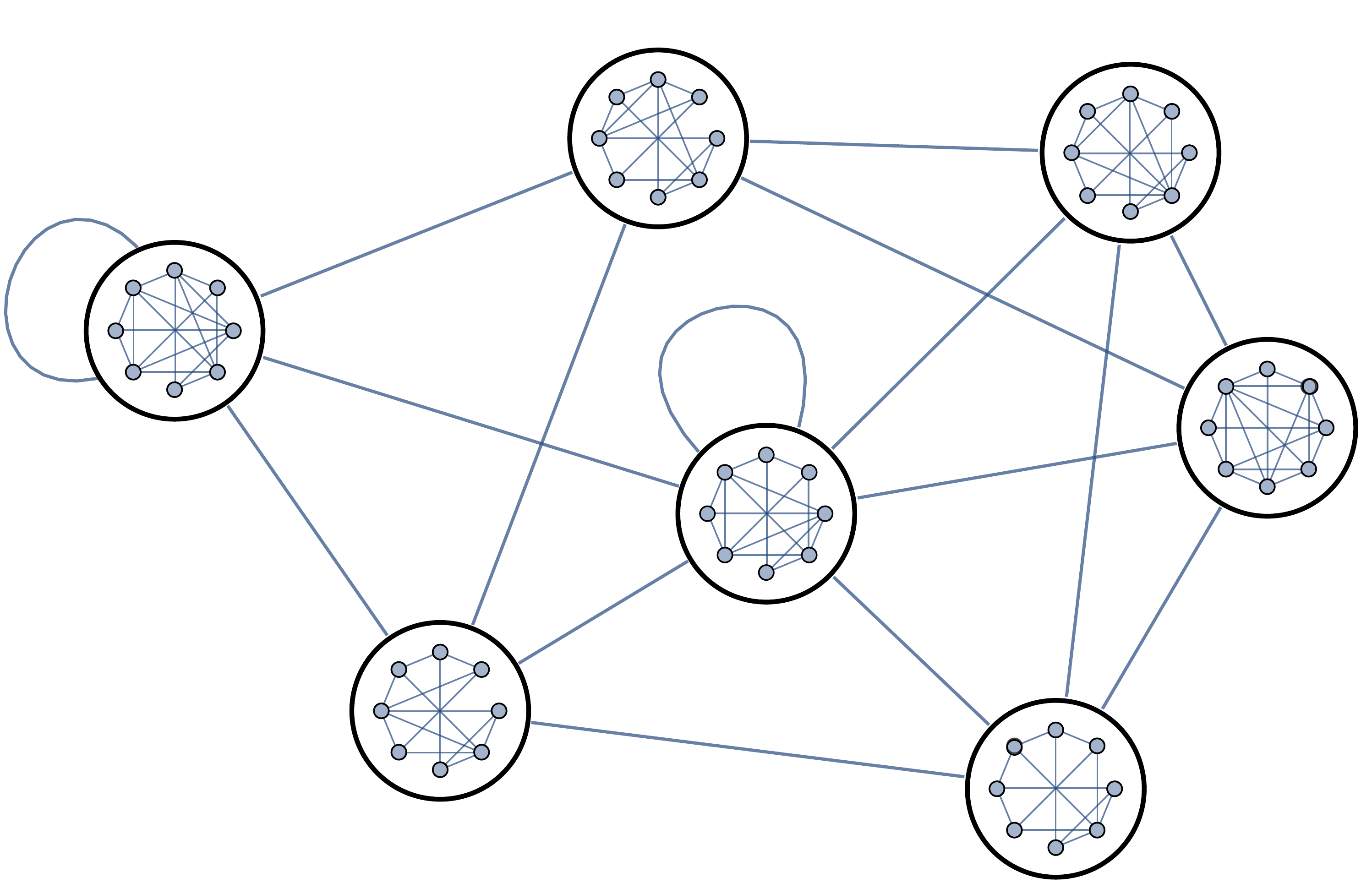}

\caption{Orbit $C_{145}$.}
\label{img:orbit145}

\end{figure*}
\clearpage

\begin{figure*}[p]
\centering

\captionsetup{width=1.0\textwidth}
\includegraphics[width=1.0\textwidth,center]{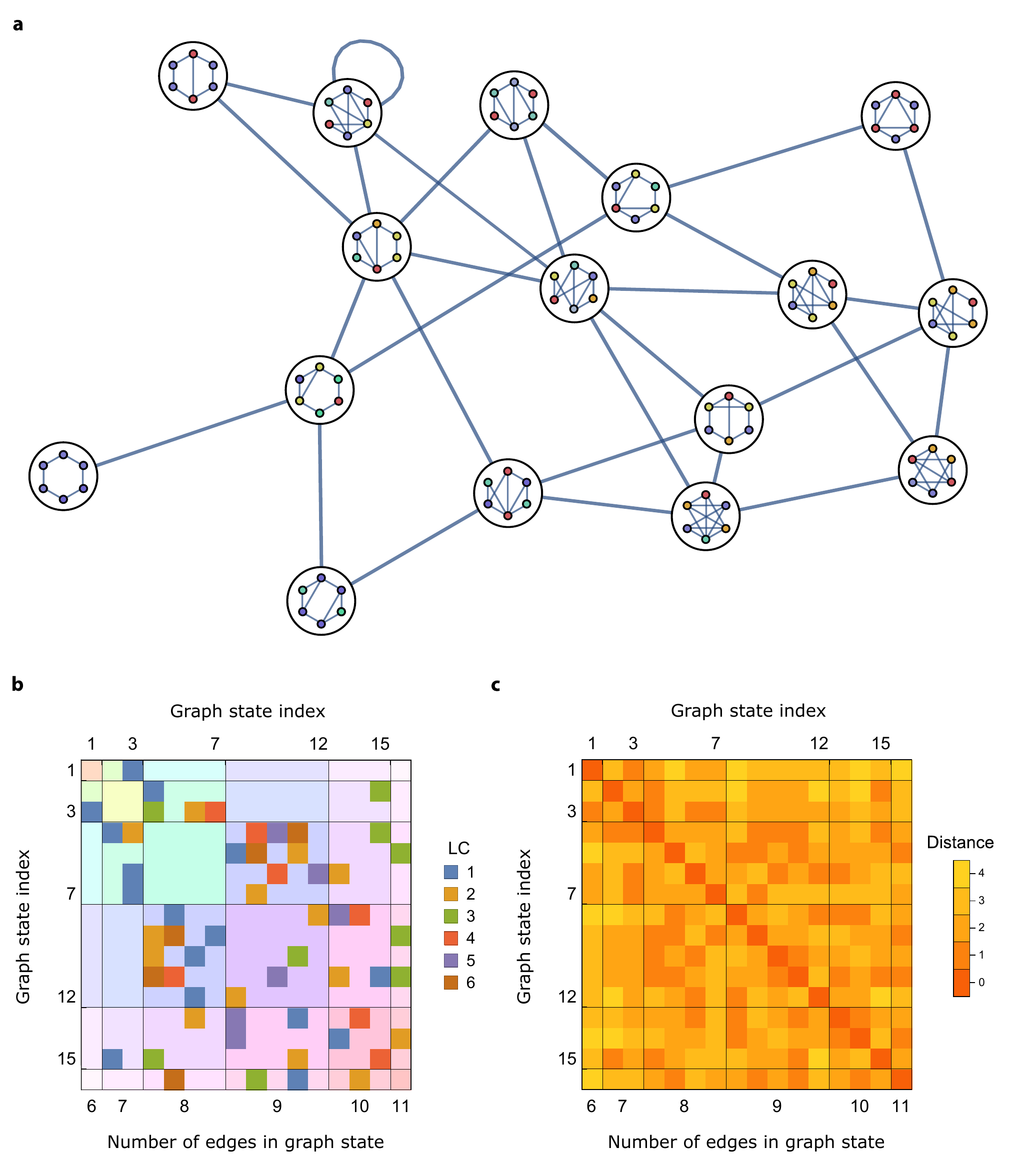}

\caption{Local complementation orbit $C_{18}$. \textbf{a.} The orbit $C_{18}$. Here, graph state vertices which produce the same output when locally complemented are the same colour (see Section \ref{sec:ge}). \textbf{b.} The adjacency matrix of $C_{18}$. \textbf{c.} The distance matrix of $C_{18}$. Regions in the plot correspond to graph states which have the same number of edges.}
\label{img:orbit18}
\end{figure*}

\begin{figure*}[p]
\centering

\captionsetup{width=1.0\textwidth}
\includegraphics[width=1.0\textwidth,center]{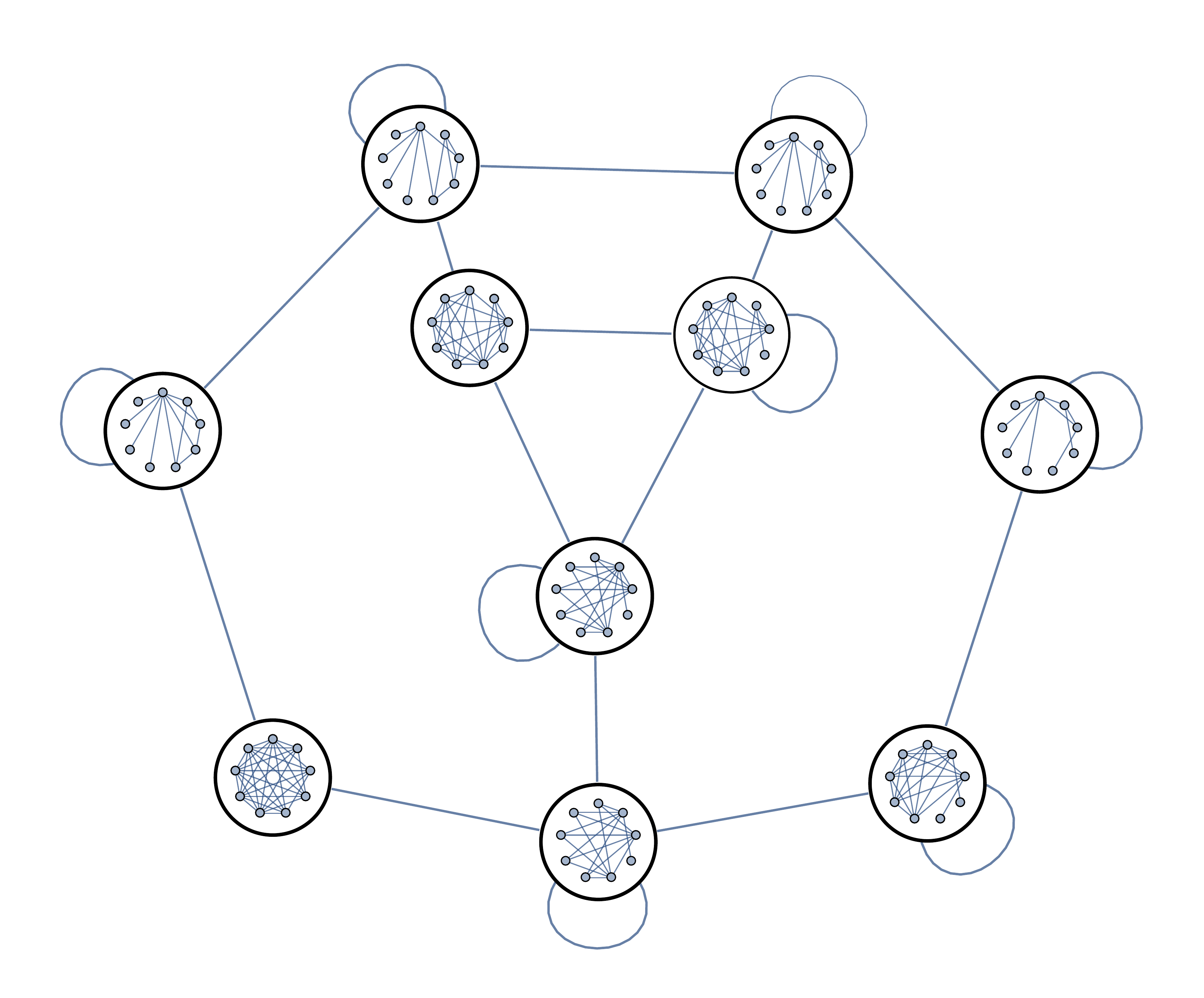}

\caption{Orbit $C_{196}$. 
}
\label{img:orbit196}
\end{figure*}

\begin{figure*}[p]
\centering

\captionsetup{width=1.0\textwidth}
\includegraphics[width=1.0\textwidth,center]{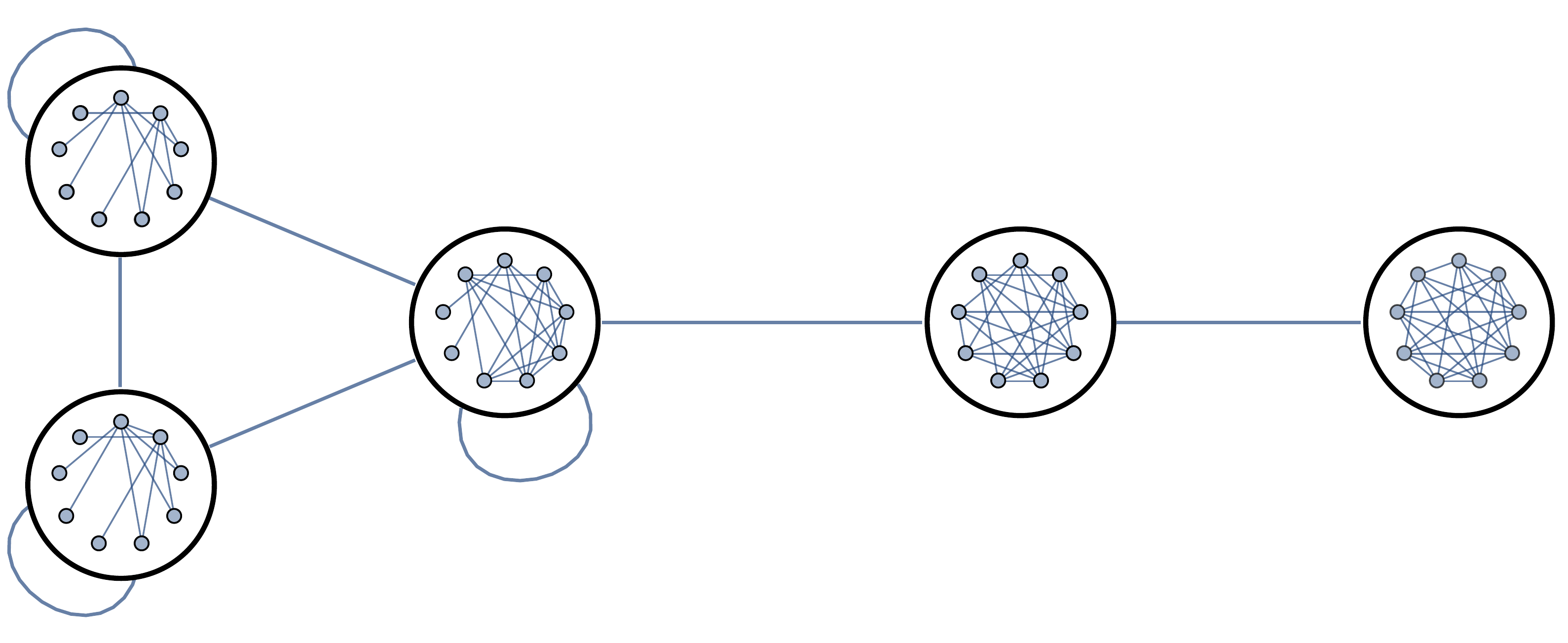}

\caption{Orbit $C_{289}$.
}
\label{img:orbit289}
\end{figure*}

\begin{figure*}[p]
\centering

\captionsetup{width=1.0\textwidth}
\includegraphics[width=1.0\textwidth,center]{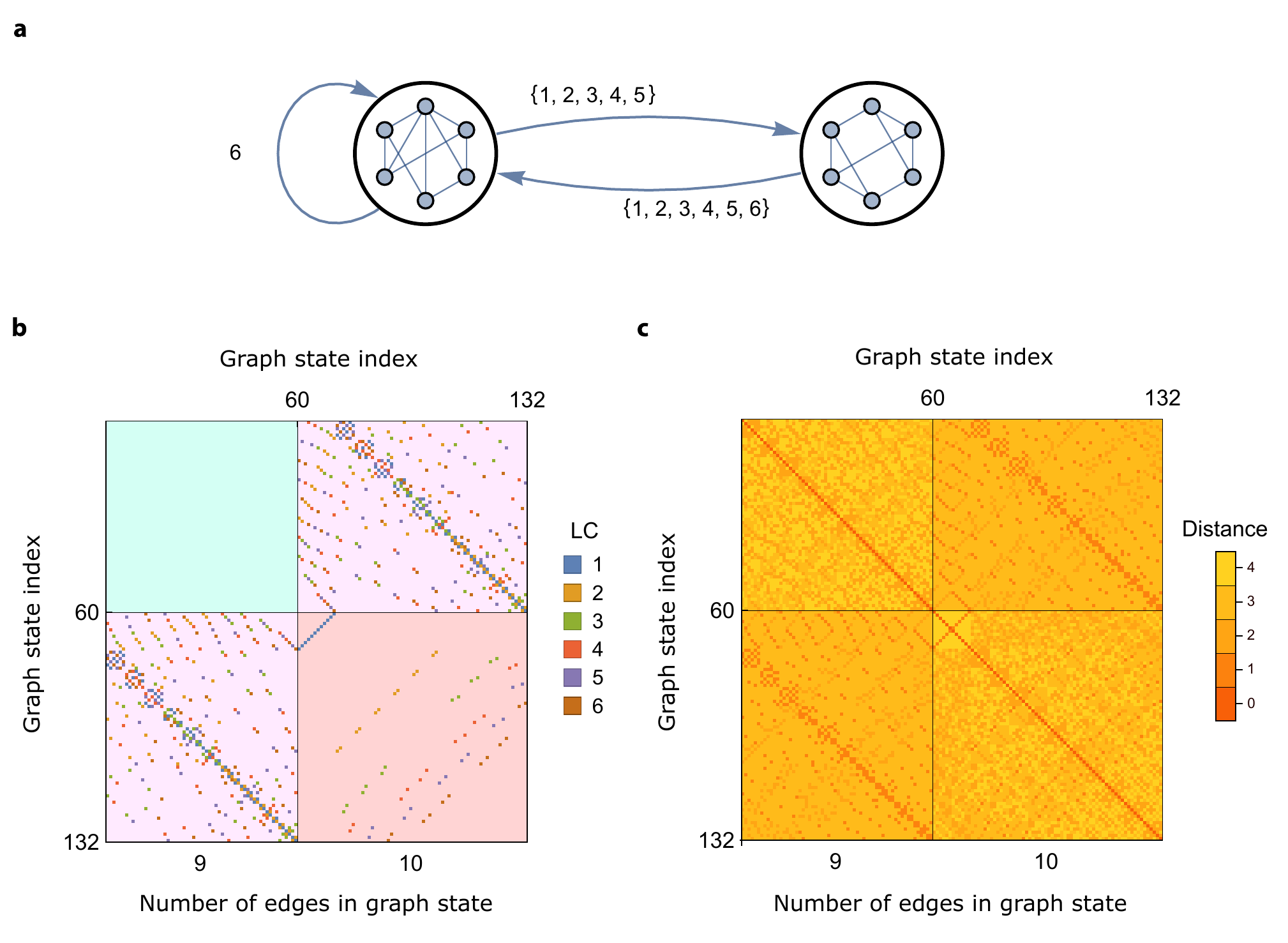}

\caption{Local complementation orbit 19. \textbf{a.} The orbit $C_{19}$. \textbf{b.} The adjacency matrix of $L_{19}$. \textbf{c.} The distance matrix of $L_{19}$. Regions in the plot correspond to graph states which are isomorphic. By separating regions corresponding to non-isomorphic graphs, we see that the orbit is composed of just two non-isomorphic graph states. These have 60 and 72 isomorphisms respectively.}

\label{img:orbit19}
\end{figure*}

\begin{figure*}[p]
\centering

\captionsetup{width=1.0\textwidth}

\includegraphics[width=1.0\textwidth,center]{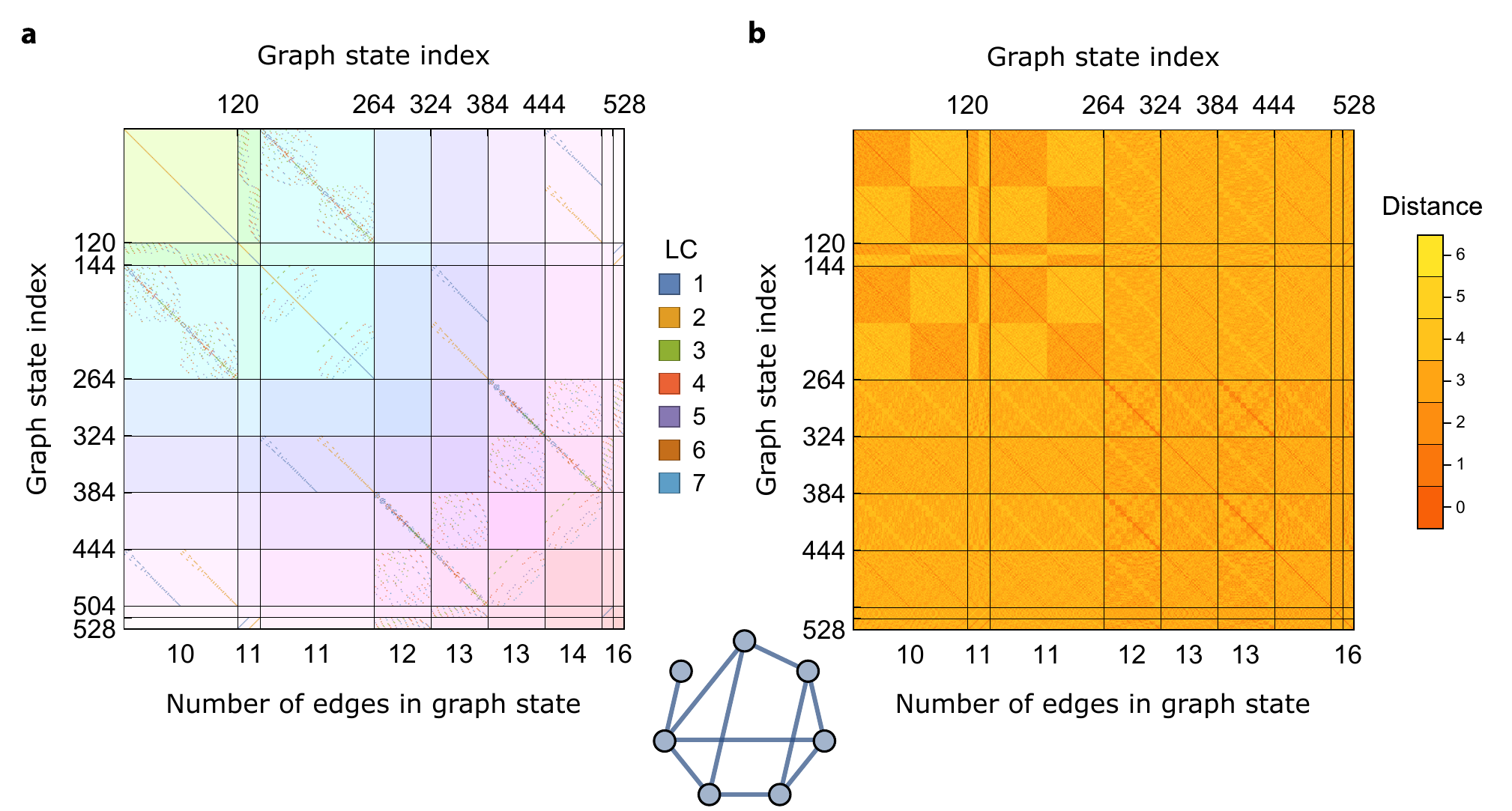}
\caption{Local complementation orbit 45. \textbf{a.} The adjacency matrix of $L_{45}$. \textbf{b.} The distance matrix of $L_{45}$.}

\label{img:adjmat-7-26}
\end{figure*}

\begin{figure*}[p]
\centering

\captionsetup{width=1.0\textwidth}
\includegraphics[width=1.0\textwidth,center]{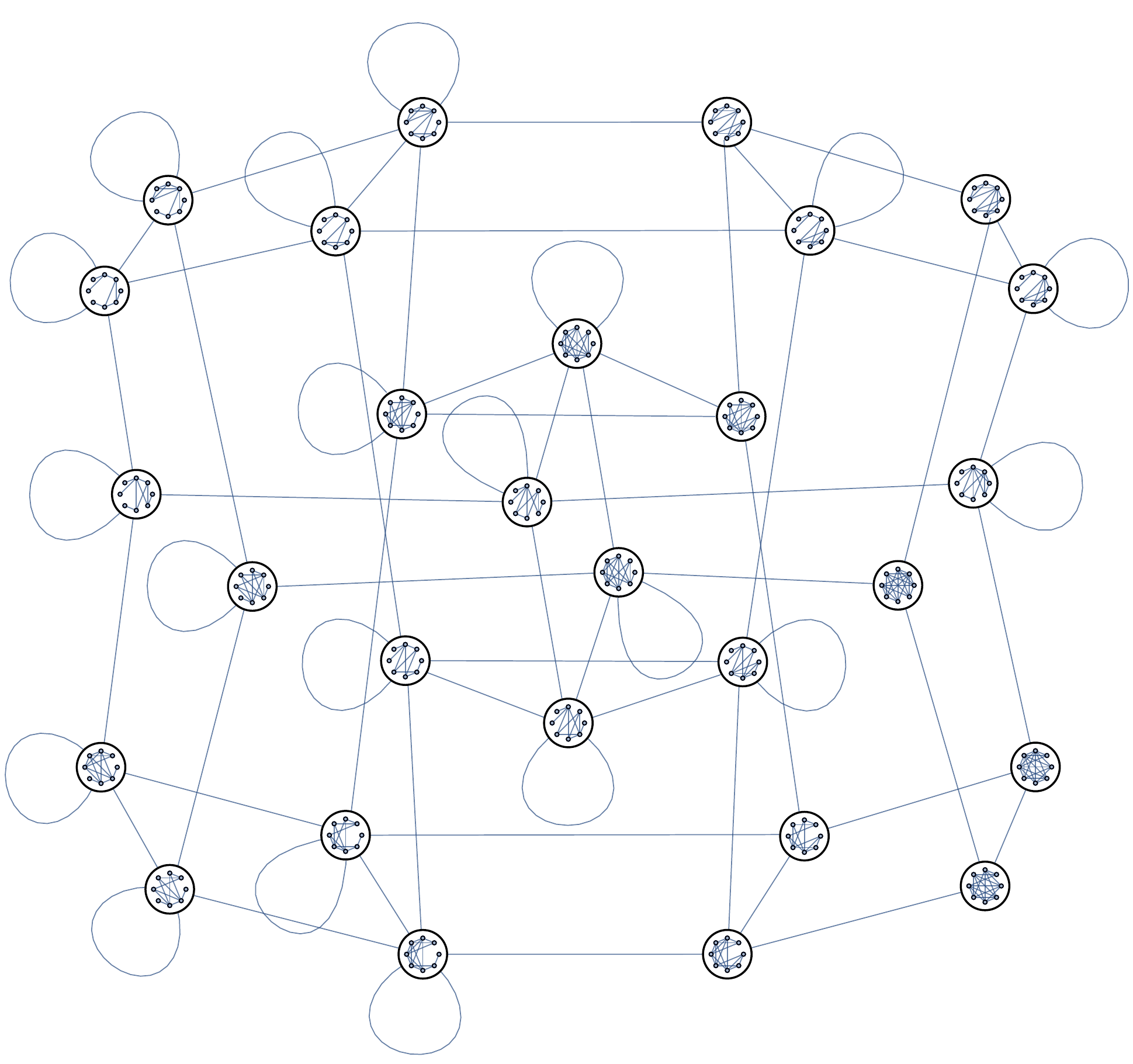}

\caption{Orbit $C_{78}$. The adjacencies and distances matrices for for $C_{78}$ and $L_{78}$ are shown on the next page. 
}

\caption{The orbit $C_{78}$.}
\label{img:orbit78}
\end{figure*}

\begin{figure*}[p]
\centering

\captionsetup{width=1.0\textwidth}
\includegraphics[width=1.0\textwidth,center]{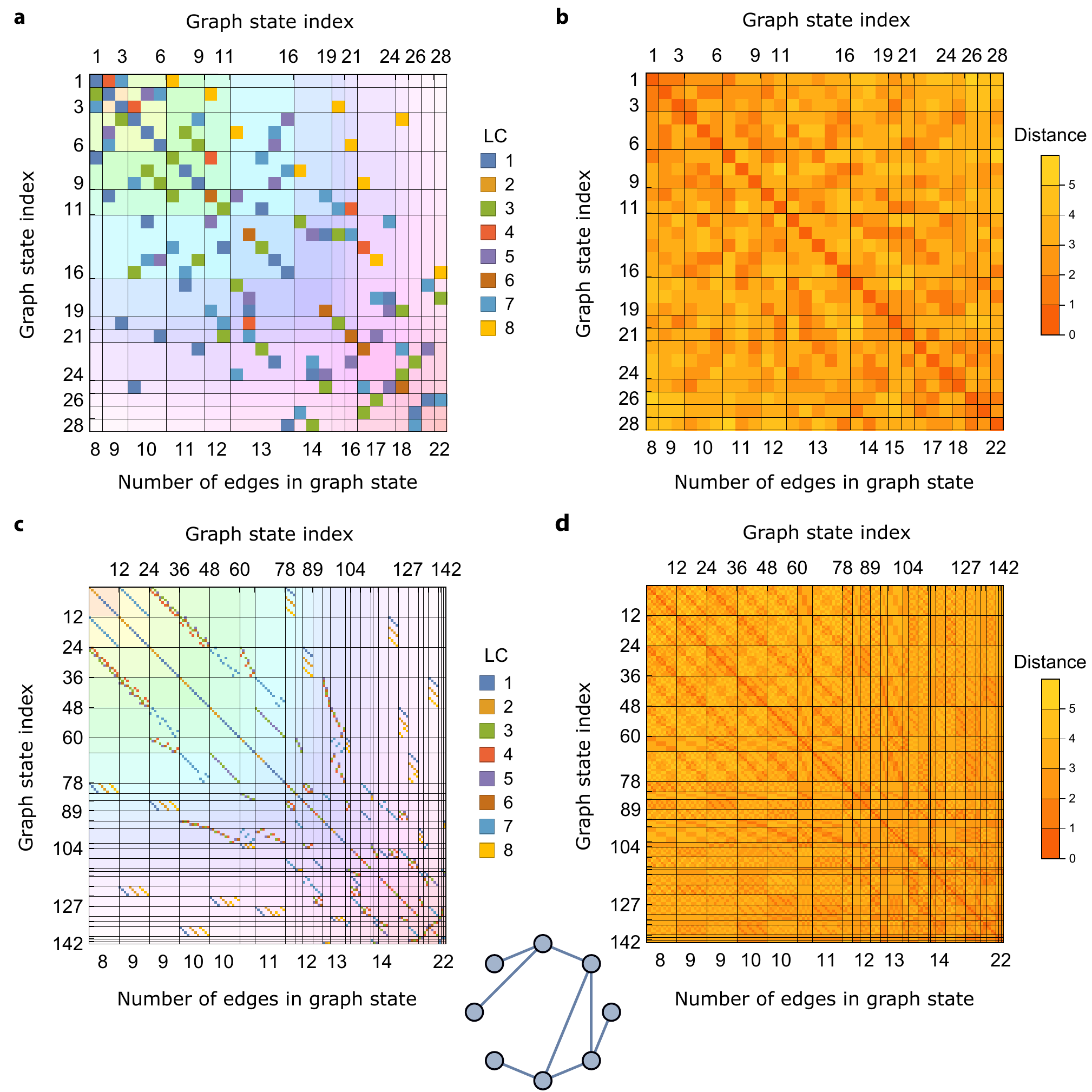}

\caption{Comparison of $L$ and $C$ orbits for enatanglement class 78. \textbf{a.} The adjacency matrix of $C_{78}$. \textbf{b.} The distance matrix of $C_{78}$. \textbf{c.} The adjacency matrix of $L_{78}$. \textbf{d.} The distance matrix of $L_{78}$. Regions in \textbf{c} and \textbf{d} correspond to graph states which have the same number of edges, while regions in \textbf{c} and \textbf{d} correspond to isomorphic graph states. Each region of \textbf{c} and \textbf{d} corresponds to a single point in \textbf{a} and \textbf{b}.}

\label{img:mat78}
\end{figure*}

\begin{figure*}[t]
\centering

\captionsetup{width=1.0\textwidth}
\includegraphics[width=1.0\textwidth,center]{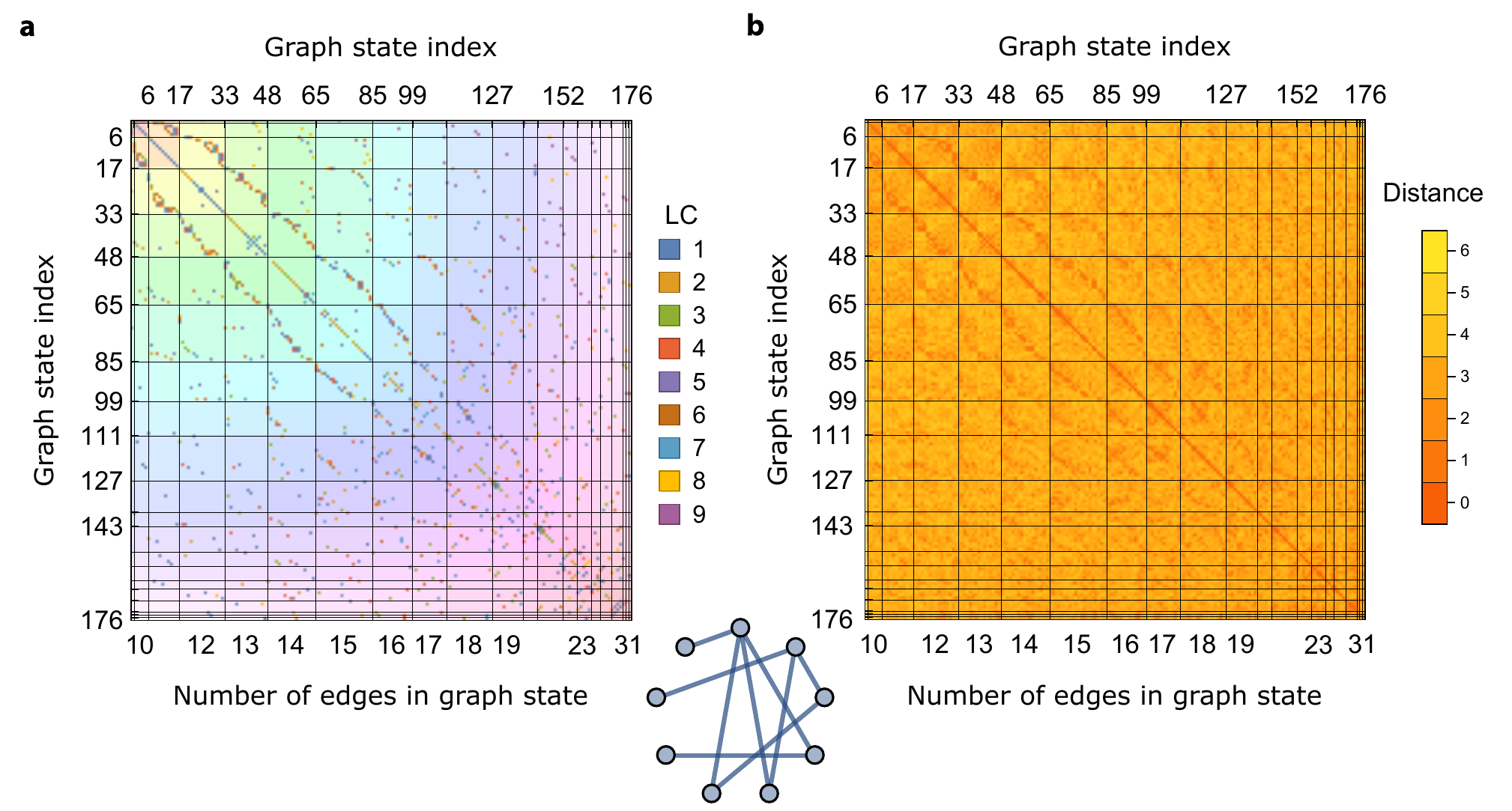}
\caption{$C_{247}$ represented by its adjacency and distance matrices. \textbf{a.} The adjacency matrix. This matrix takes a value $\Gamma_{ij}=n$ where $n$ is the index of the local complementation that links orbit vertices $i$ and $j$. A different colour is used for each $n$. We demarcate regions of the plot which correspond to graph states with the same number of edges, and shade these in the same pastel shade. \textbf{b.} The distance matrix of $C_{247}$. This matrix takes a value $\Gamma_{ij}=n$ where $n$ is the number of local complementations needed to transform graph state $i$ into graph state $j$. The canonical representative graph state of the class is shown underneath. }

\label{img:adjmat-9-101}
\end{figure*}

\clearpage
\section{Representive members of all graph state entanglement classes of  $n<9$ qubits}
\label{sec:allstates}

\begin{figure*}[!ht]
\centering
\captionsetup{width=1.0\textwidth}
\includegraphics[width=1.0\textwidth]{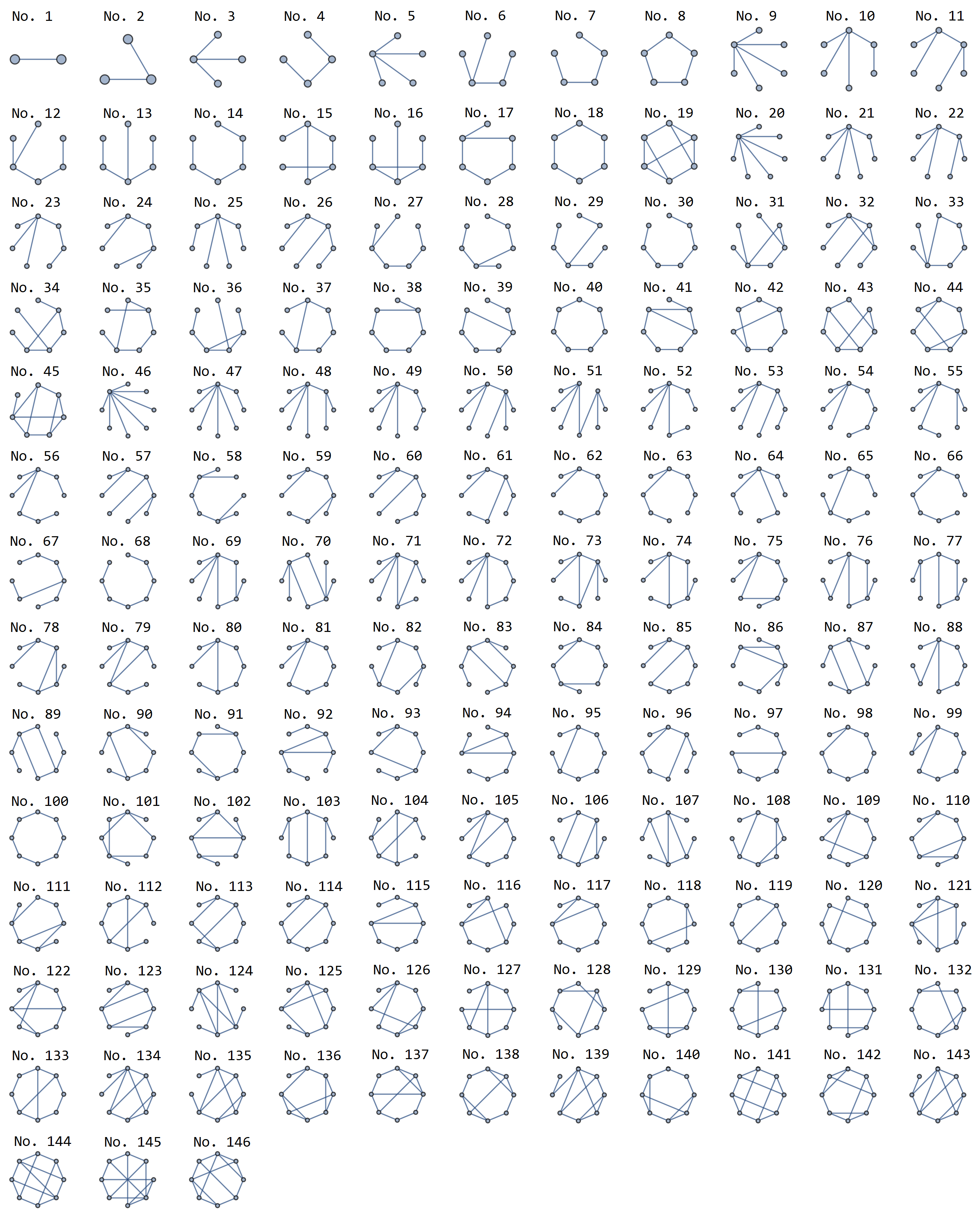}

\caption{Canonical, minimal edge count representatives from each orbit up to and including 8 qubits. These are canonically indexed as in ref.~\citenum{cabello2011optimal}.}
\label{img:bigtable}
\end{figure*}

\end{document}